\documentclass[%
 reprint,
 amsmath,amssymb,
 aps, onecolumn
]{revtex4-2}

\usepackage{xcolor}
\usepackage{graphicx}% Include figure files
\usepackage{subcaption}
\usepackage{dcolumn}% Align table columns on decimal point
\usepackage{bm}
\usepackage{color}
\usepackage{float}
\usepackage{subcaption}
\usepackage{ragged2e}
\usepackage[unicode=true,pdfusetitle,
 bookmarks=true,bookmarksnumbered=false,bookmarksopen=false,
 breaklinks=false,pdfborder={0 0 1},backref=false,colorlinks=false]
 {hyperref}
 \hypersetup{colorlinks,allcolors=blue,breaklinks=true}
\usepackage{natbib}
\usepackage{bibunits}
\defaultbibliographystyle{apsrev4-2}
\newcounter{bibsectioncount}

\begin{document}

\let\savedaddcontentsline\addcontentsline

\renewcommand{\addcontentsline}[3]{}
\stepcounter{bibsectioncount} 
\begin{bibunit}

\title{Observation of Berry curvature fluctuations from incipient polar order in an oxide interface
}
\author{S. Avraham}
\affiliation{School of Physics and Astronomy, Tel Aviv University, Tel Aviv 6997801, Israel}

\author{D. Gitman}
\affiliation{School of Physics and Astronomy, Tel Aviv University, Tel Aviv 6997801, Israel}

\author{P. Matus}
\affiliation{School of Physics and Astronomy, Tel Aviv University, Tel Aviv 6997801, Israel}

\author{S. Sandik}
\affiliation{School of Physics and Astronomy, Tel Aviv University, Tel Aviv 6997801, Israel}

\author{E. Raz}
\affiliation{School of Physics and Astronomy, Tel Aviv University, Tel Aviv 6997801, Israel}

\author{S. Jana}
\affiliation{School of Physics and Astronomy, Tel Aviv University, Tel Aviv 6997801, Israel}

\author{M. Dahan}
\affiliation{School of Physics and Astronomy, Tel Aviv University, Tel Aviv 6997801, Israel}

\author{T. Holder}
\email{tobiasholder@tauex.tau.ac.il}
\affiliation{School of Physics and Astronomy, Tel Aviv University, Tel Aviv 6997801, Israel}

\author{Y. Dagan}
\email{yodagan@tauex.tau.ac.il}
\affiliation{School of Physics and Astronomy, Tel Aviv University, Tel Aviv 6997801, Israel}

\date{\today}
\begin{abstract}
Diagnosing hidden local orders at buried interfaces remains a central challenge in the design and characterization of quantum materials. Second-order electrical responses, such as the nonlinear Hall effect, probe inversion-symmetry-breaking terms invisible to linear transport, offering a direct window into these nanoscale environments via the quantum geometry of Bloch electrons. Here, we utilize $\text{KTaO}_3$, a complex oxide driven by strong tantalum $5d$ spin-orbit coupling and interfacial inversion symmetry breaking, to demonstrate that second-harmonic resistivities exhibit large, reproducible mesoscopic fluctuations. Remarkably, these fluctuations persist in macroscopically large ($200\,\mu\text{m}$) devices and are ubiquitous across all studied surface orientations, even where macroscopic conductivity strictly adheres to underlying crystal symmetries. We propose that these robust, magnetic-field-driven interference patterns arise from local structural symmetry breaking, driven by incipient ferroelectric polarization pinned to the interfacial impurity landscape. This defect-pinned polar mechanism is firmly supported by the signal's suppression above $10\text{ K}$ due to phase decoherence, and a complete loss of mesoscopic memory upon thermal cycling above $40\text{ K}$. By linking quantum geometry to dynamic lattice ordering, our findings establish nonlinear mesoscopic transport as a powerful new characterization tool, capable of revealing local polar tendencies and hidden structural orders in complex materials that remain fundamentally invisible to conventional probes.
\end{abstract}
    \maketitle
Electrical transport in time-reversal invariant crystalline solids is most often discussed within the framework of linear response, with the current expressed as a sum of the conventional Drude terms. Beyond the linear regime, however, new contributions emerge that are dictated by the quantum geometry of Bloch wavefunctions and can be symmetry-forbidden at first order \cite{
% moore2010confinement, 
du2021nonlinear, Ma2021,2025RPPhJiang}. In particular, in systems where time-reversal symmetry eliminates the linear anomalous Hall effect, second-order responses provide direct access to geometric tensors of the Bloch manifold: the Berry curvature, which acts as an effective magnetic field in momentum space, and the quantum metric, which quantifies distances between nearby quantum states in Hilbert space \cite{
gao2019nonreciprocal,lapa2019semiclassical,
yu2025quantum,verma2026quantum}. While most theoretical treatments express these geometric contributions as bulk averages of pristine crystals, how they are reshaped by local disorder, and how these disorder-sensitive observables are tuned by magnetic field and chemical potential, remain open questions~\cite{Corbae2023,Gao2025review}.
\par
A prominent example of such a geometric phenomenon is the nonlinear Hall effect, which emerges as a dissipative second-harmonic ($2\omega$) response to an applied alternating current of frequency $\omega$ \cite{Shi2023}, strictly enabled by the breaking of inversion symmetry ~\cite{Shi2023}. When intrinsic mechanisms dominate, this signal can be traced to the Berry curvature dipole (BCD), the first moment of Berry curvature over the Fermi surface \cite{sodemann2015quantum,ma2019observation,du2021nonlinear}. Physically, an oscillating field drives the nonequilibrium carrier distribution to momentum regions of opposite Berry curvature asymmetrically, generating a transverse current proportional to the BCD; concurrently, extrinsic mechanisms (e.g., skew scattering) can also yield large nonlinear Hall signals \cite{du2021quantum, he2021quantum}.
\par
Potassium tantalate, $\mathrm{KTaO_3}$ (KTO), is a paradigmatic quantum paraelectric (incipient ferroelectric) whose long-range ferroelectric order is suppressed to the lowest temperatures by quantum fluctuations, closely paralleling the canonical behavior of $\mathrm{SrTiO_3}$ \cite{Rowley2014,MullerBurkard1979}. Crucially, however, KTO hosts Ta $5d$-derived conduction bands with significantly larger atomic spin-orbit coupling than that found in Ti $3d$ titanates, strongly reshaping the band topology near the Fermi contour and enabling pronounced spin-orbital textures under confinement and interfacial inversion breaking \cite{Bruno2019,King2012}. This combination of quantum-paraelectric lattice polarizability and strong spin-orbit coupling makes KTO-based 2DES a fertile arena where quantum geometry can be directly linked to measurable nonlinear transport. Previous nonlinear transport experiments on the related titanate $\mathrm{SrTiO_3}$ have revealed tunable Dirac cones in Rashba-split bands \cite{tuvia2024enhanced} and behavior associated with the quantum metric of the system \cite{sala2025quantum}.
\par
High-mobility 2D electron systems realized at KTO surfaces and interfaces exhibit a striking crystal-face dependence: (100)-oriented interfaces can remain nonsuperconducting to ultralow temperatures, whereas 2D electron systems realized at (111) and (110) oriented surfaces have shown substantially higher-$T_c$ superconductivity and pronounced anisotropies, underscoring the decisive role of confinement direction and symmetry \cite{LiuScience2021,RenSciAdv2022,NormanNatCommun2023}. The same symmetry distinctions shape nonlinear responses: for example, large nonlinear Hall signals have been reported in KTO (111)-based 2DES and analyzed in terms of dominant skew scattering with gate-tunable features tied to Berry-curvature hot-spots near avoided crossings \cite{ZhaiNanoLett2023}. 
On the other hand, even for the nominally symmetric (100) and (110) orientations, different microscopic facets may emerge that relax the macroscopic symmetry constraints, allowing otherwise forbidden nonlinear transport terms to appear; for the (110) surface, this is often driven by local surface reconstructions \cite{WangPCCP2018,Martinez2023}.
\par 
Here, we report a systematic investigation of nonlinear transport across various $\mathrm{LaAlO_3/KTaO_3}$ (LAO/KTO) interfaces, revealing that the second-order response is marked by conspicuous, reproducible fluctuation patterns, whose strong sensitivity to an out-of-plane magnetic field provides direct evidence of quantum interference.
\par
\begin{figure*}
\begin{subcaptiongroup}
\includegraphics[width=.9\columnwidth]{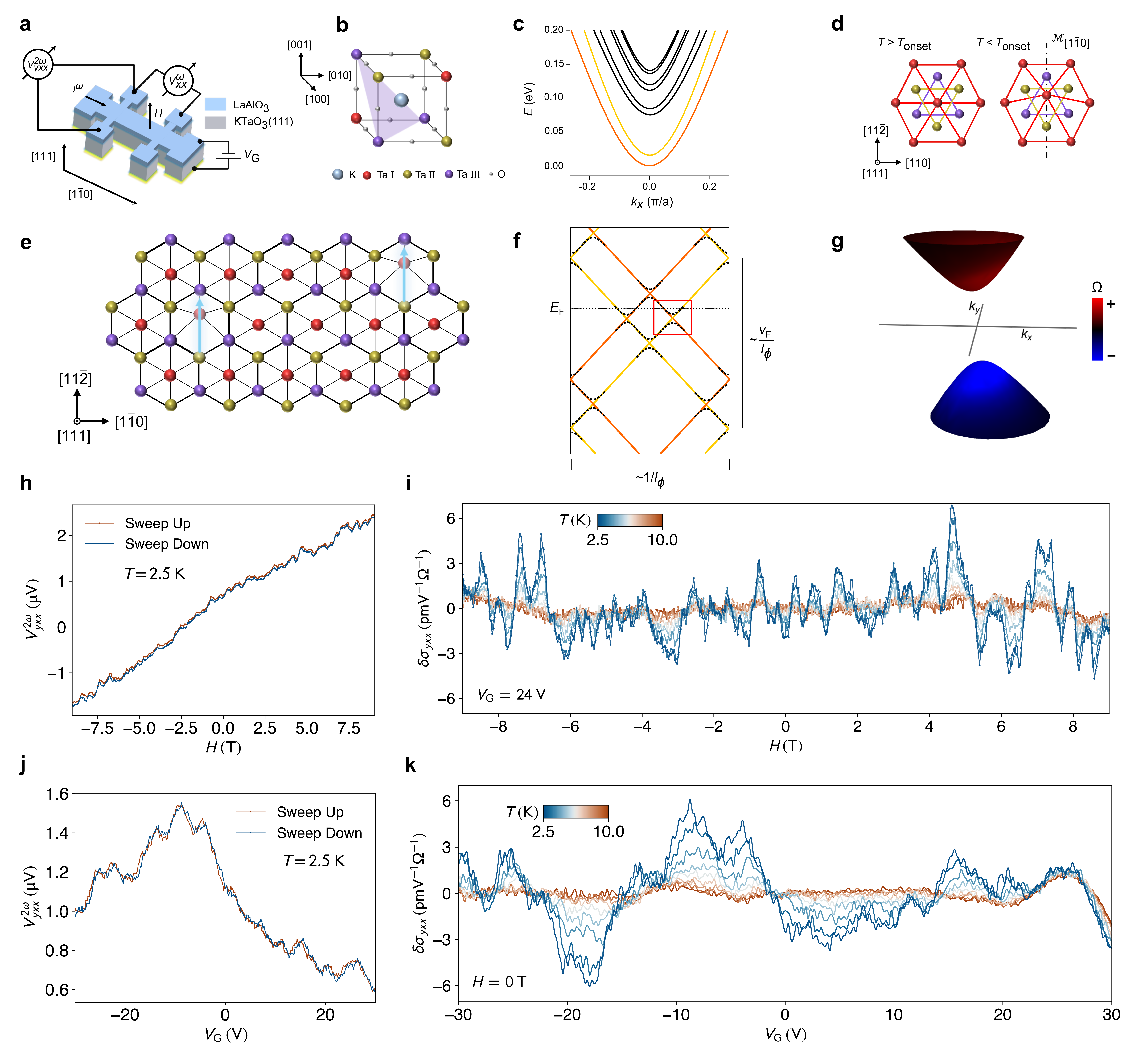}% Here is how to import EPS art
\phantomcaption\label{fig:summary_a}
\phantomcaption\label{fig:summary_b}
\phantomcaption\label{fig:summary_c}
\phantomcaption\label{fig:summary_d}
\phantomcaption\label{fig:summary_e}
\phantomcaption\label{fig:summary_f}
\phantomcaption\label{fig:summary_g}
\phantomcaption\label{fig:summary_h}
\phantomcaption\label{fig:summary_i}
\phantomcaption\label{fig:summary_j}
\phantomcaption\label{fig:summary_k}
\end{subcaptiongroup}
\captionsetup{subrefformat=parens}
\caption{\label{fig:summary}  \justifying 
\textbf{Berry curvature fluctuations in KTaO$_3$(111) two-dimensional electron system}. \subref*{fig:summary_a} Scheme of the measured $\mathrm{LaAlO_3/KTaO_3(111)}$ interface 2DES Hall bar device. An alternating current at frequency $\omega$ and amplitude $I^\omega$ is applied. The linear longitudinal voltage $V_{xx}^\omega$ and the second-harmonic nonlinear Hall voltage $V_{yxx}^{2\omega}$ are measured. \subref*
{fig:summary_b} Schematic of the unit cell of the quantum paraelectric KTaO$_3$. The (111) plane is indicated. \subref*{fig:summary_c} Schematic energy-momentum dispersion along the $[1\bar{1}0]$ direction of the two-dimensional electron system at the KTaO$_3$(111) surface (based on Ref.~\cite{Bruno2019}). The outermost pair of bands is indicated. \subref*{fig:summary_d} Surface symmetry breaking. At high temperatures, the KTaO$_3$(111) oriented surface exhibits three-fold rotational symmetry. Below an onset temperature, $T_\text{onset}$, spontaneous surface symmetry breaking occurs. The diagram illustrates a displacement of the central Ta atom, which limits the system's symmetry to a single mirror plane. \subref*{fig:summary_e} Mesoscopic illustration of the KTaO$_3$(111) surface disordered by local ferroelectric distortions at low temperatures. Spontaneous rotational symmetry breaking is manifested as the creation of local electric dipoles (light blue arrows), which create a disordered ferroelectric state. \subref*{fig:summary_f} Energy-momentum dispersion of a mesoscopic supercell of linear size $l_\phi$ (area $\sim l_\phi^2$) along the $[1\bar{1}0]$ direction. Only the folded bands originating from the outermost pair of bands in \subref*{fig:summary_c} are shown. The presence of a local disorder site, which pins the ferroelectric polarization, is manifested as a tilting of the energy bands and the opening of a gap at the Dirac points (massive tilted Dirac nodes). A red box highlights a tilted gap at a Dirac point near the Fermi surface.  
\subref*{fig:summary_g} Tilted massive Dirac node. The resulting Berry curvature distribution is asymmetric, leading to a finite Berry curvature dipole pointing perpendicularly to the mirror plane. \subref*{fig:summary_h} 
Raw second-harmonic Hall voltage, $V_{yxx}^{2\omega}$, as a function of an out-of-plane magnetic field $H$. Plots for opposite field sweep directions (red and blue curves) are presented, showing reproducible fluctuations. \subref*{fig:summary_i} Extracted fluctuations in the second-harmonic Hall conductivity, $\delta \sigma_{yxx}$, as a function of an out-of-plane magnetic field $H$ at various temperatures $T$. \subref*{fig:summary_j} Raw second-harmonic Hall voltage, $V_{yxx}^{2\omega}$, as a function of back-gate voltage $V_\mathrm{G}$ for opposite gate voltage sweep directions (red and blue curves). Conspicuous and reproducible fluctuations are observed. \subref*{fig:summary_k} Extracted $\delta \sigma_{yxx}$ as a function of $V_\mathrm{G}$ at various temperatures $T$.}
\end{figure*}

To demonstrate the properties of the nonlinear conductance fluctuation patterns, we focus first on the (111) interface as a primary example (see Methods and Fig. \ref{fig:summary_a} for experimental setup scheme). The $\mathrm{KTaO_3}$ unit cell and the indication of the (111) plane are shown in Fig. \ref{fig:summary_b}. The schematic energy-momentum dispersion for this interface, featuring Rashba spin-orbit split bands, is presented in Fig. \ref{fig:summary_c}. Fig. \ref{fig:summary_d} illustrates how a spontaneous shift of surface Ta reduces the pristine surface symmetry to a single mirror line. Figs. \ref{fig:summary_e} through \ref{fig:summary_g} further outline how local disorder sites pin this ferroelectric polarization, creating massive, tilted Dirac nodes that generate an asymmetric Berry curvature distribution and a finite, disordered Berry curvature dipole landscape.
\par
Our main results are shown in Figs. \ref{fig:summary_h} through \ref{fig:summary_k}. Raw data for the Second-harmonic Hall voltage, $V_{yxx}^{2\omega}$, as a function of magnetic field are presented in Fig. \ref{fig:summary_h}. A clear fluctuation pattern appears as fine-scale deviations superimposed on a monotonic background trend. The high reproducibility of these features is evident by comparing the two consecutive magnetic field sweeps. To isolate these fluctuations, we subtract a smooth, monotonic background, $\bar{V}_{yxx}^{2\omega}$, yielding the voltage fluctuations $\delta V_{yxx}^{2\omega}(H)=V_{yxx}^{2\omega}(H)-\bar{V}_{yxx}^{2\omega}(H)$. We subsequently convert these to nonlinear Hall conductivity fluctuations, $\delta\sigma_{yxx}=\frac{\sqrt{2}L^3}{W^2}\frac{I^\omega \delta V_{yxx}^{2\omega}}{(V_{xx}^{\omega})^3}$, where $L$ and $W$ are the length and width of the device, respectively, and $I^{\omega}$ is the drive current (see Supplementary Sec. II for further details).
\par
We find that these fluctuations are highly sensitive to external tuning parameters. They vary sharply with the out-of-plane magnetic field (Fig. \ref{fig:summary_i}). Furthermore, they evolve systematically with gate voltage (see raw data in Fig. \ref{fig:summary_j}), which tunes both the chemical potential and the Rashba spin-orbit coupling \cite{PhysRevLett.104.126802}. As demonstrated in Figs. \ref{fig:summary_i} and \ref{fig:summary_k}, this fluctuating signal is visible up to a characteristic temperature of $T \simeq 10$ K. Notably, these features are ubiquitous across all three KTO surface orientations studied (see also Figs. \ref{fig:fig3_f}, \ref{fig:fig3_g}).
\par

\par
\begin{figure*}
\begin{subcaptiongroup}
\includegraphics[width=1\columnwidth]{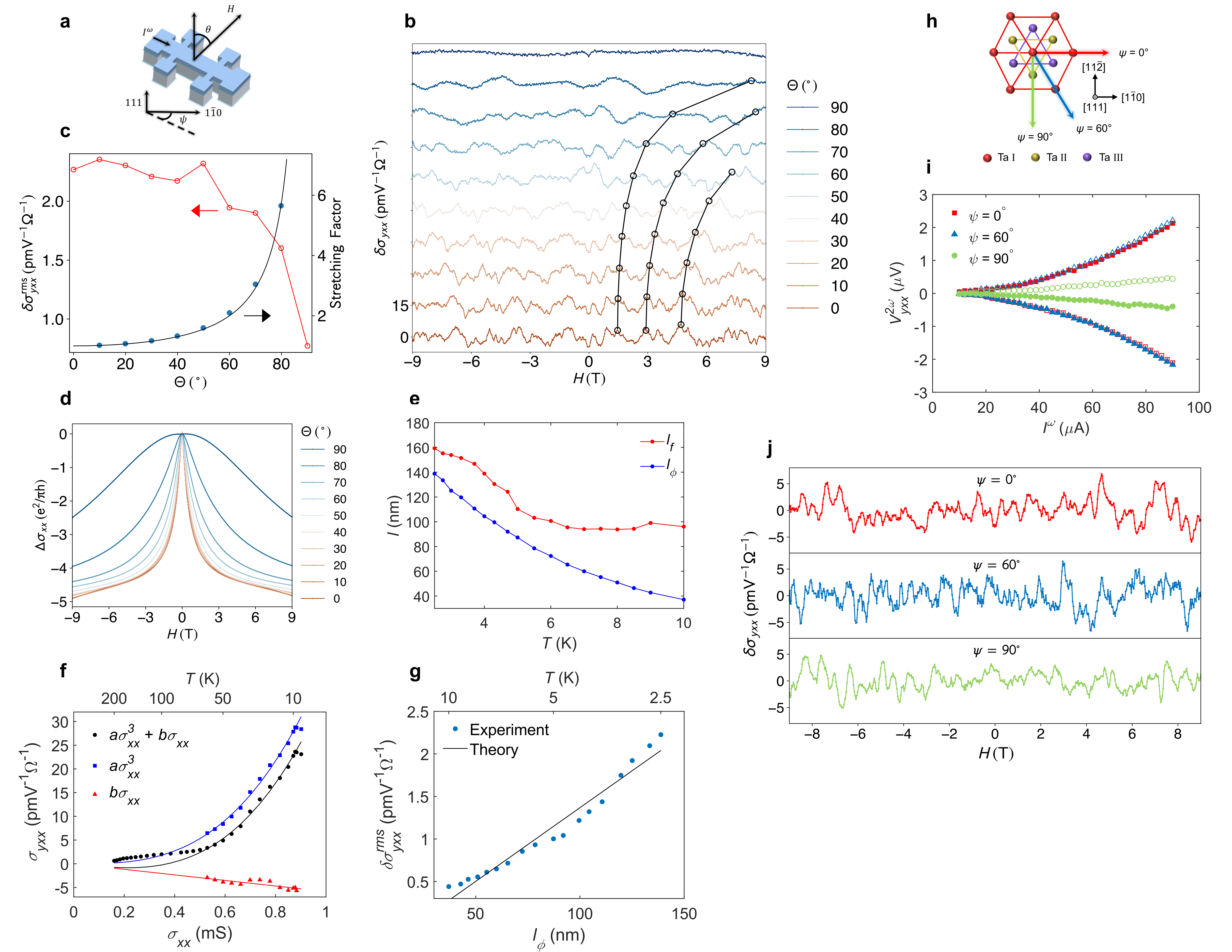}% Here is how to import EPS art
\phantomcaption\label{fig:fig2_a}
\phantomcaption\label{fig:fig2_b}
\phantomcaption\label{fig:fig2_c}
\phantomcaption\label{fig:fig2_d}
\phantomcaption\label{fig:fig2_e}
\phantomcaption\label{fig:fig2_f}
\phantomcaption\label{fig:fig2_g}
\phantomcaption\label{fig:fig2_h}
\phantomcaption\label{fig:fig2_i}
\phantomcaption\label{fig:fig2_j}
\end{subcaptiongroup}
\captionsetup{subrefformat=parens}
\caption{\label{fig:fig2} \justifying \textbf{Angular and temperature dependence of Berry curvature fluctuations.} \subref*{fig:fig2_a} Sketch of the measurement configuration. We define $\theta$ as the sample tilt angle with respect to the magnetic field, and $\psi$ as the angle between the driving current and the $[1\bar{1}0]$ axis. The plane of rotation is chosen such that the resulting in-plane field component is collinear with the current. \subref*{fig:fig2_b} Nonlinear Hall fluctuations, $\delta \sigma_{yxx}$, as a function of magnetic field strength $H$ for selected tilt angles $\theta$. Consecutive curves are vertically offset by $15\ \mathrm{pmV^{-1}\Omega^{-1}}$ for clarity. \subref*{fig:fig2_c} Stretching factor and root-mean-square amplitude of the fluctuations, $\delta \sigma_{yxx}^\text{rms}$, as a function of tilt angle $\theta$. The stretching factor follows $1/\cos\theta$ dependence, consistent with the perpendicular field component (black solid line). The amplitude $\delta \sigma_{yxx}^\text{rms}$ decays with increasing $\theta$ (red solid line).  \subref*{fig:fig2_d} The longitudinal magneto-conductivity $\Delta \sigma_{xx}(H)=\sigma_{xx}(H)-\sigma_{xx}(0)$ as a function of $H$ at selected tilt angles. \subref*{fig:fig2_e} Extracted length scales at $\theta=0^{\circ}$ as a function of temperature $T$: the phase coherence length $l_\phi$ extracted from the WAL fitting, and the characteristic fluctuation length $l_f$ extracted from the nonlinear fluctuation patterns (see main text for discussion). \subref*{fig:fig2_f} Macroscopic nonlinear Hall conductivity $\sigma_{yxx}$ as a function of longitudinal conductivity $\sigma_{xx}$ (black dots). The conductivity is tuned by varying the temperature (top horizontal axis). The low-temperature data ($T < 60$ K) are fitted to $a\sigma_{xx}^{3} + b\sigma_{xx}$ dependence (black solid line). The total $\sigma_{yxx}$ is separated into a dominant skew scattering contribution ($\sim a\sigma_{xx}^3$, blue squares) and a smaller contribution ($\sim b\sigma_{xx}$, red triangles), which may originate from the Berry curvature dipole. We extract fitting parameters $a=42.3\pm2.6\ \mathrm{mm\,V^{-1}\Omega^{2}}$ and $b=-5.9\pm1.5\ \mathrm{nm\,V^{-1}}$. \subref*{fig:fig2_g} Fluctuation amplitude $\delta \sigma_{yxx}^\text{rms}$ as a function of phase coherence length $l_\phi$, tuned by temperature (top horizontal axis). The solid line represents a linear fit consistent with our model. We obtain $\delta\sigma_{yxx}^{rms}=ml_\phi+c$, where $m=17.3\pm1.9\ \mathrm{\mu SV^{-1}}$ and $c=-0.36\pm0.17\ \mathrm{pmV^{-1}\Omega^{-1}}$. \subref*{fig:fig2_h} Schematic of three selected Hall-bar orientations with respect to the crystal axis: $\psi=0^{\circ}$, $\psi=60^{\circ}$, and $\psi=90^{\circ}$ within the (111) plane. \subref*{fig:fig2_i} The second-harmonic Hall voltage $V_{yxx}^{2\omega}$ as a function of AC current amplitude $I^{\omega}$, measured at the selected Hall-bar orientations $\psi$. Filled and open markers indicate positive and negative current directions, respectively, measured by swapping the voltage probes. \subref*{fig:fig2_j} Fluctuations $\delta \sigma_{yxx}$ as a function of out-of-plane magnetic field $H$, measured at the selected angles $\psi$. While the macroscopic conductivity follows the crystal symmetry, the fluctuation patterns exhibit similar magnitudes but distinct mesoscopic features.}
\end{figure*}
\begin{figure*}
\begin{subcaptiongroup}
\includegraphics[width=1\columnwidth]{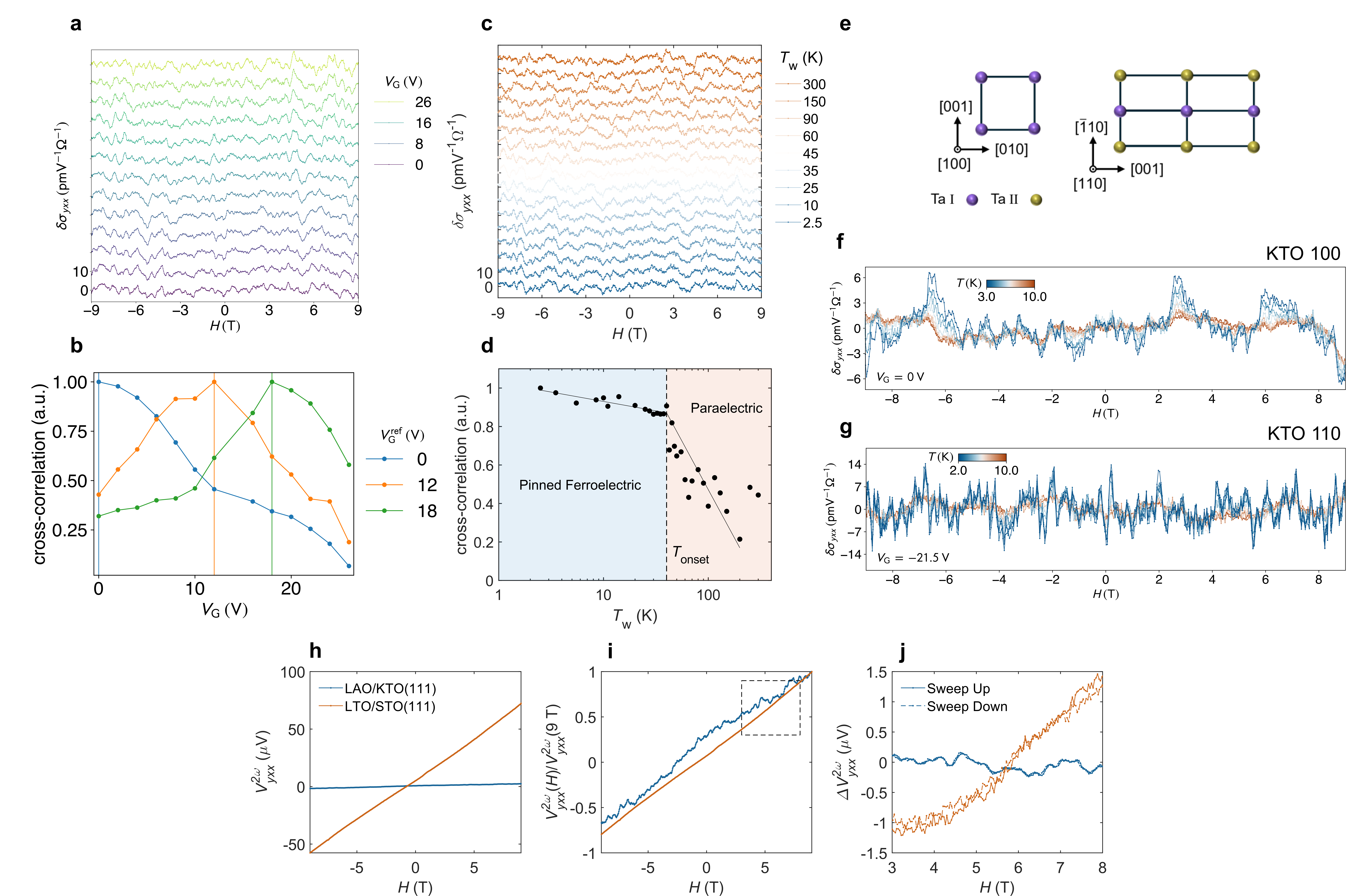}% Here is how to import EPS art
\phantomcaption\label{fig:fig3_a}
\phantomcaption\label{fig:fig3_b}
\phantomcaption\label{fig:fig3_c}
\phantomcaption\label{fig:fig3_d}
\phantomcaption\label{fig:fig3_e}
\phantomcaption\label{fig:fig3_f}
\phantomcaption\label{fig:fig3_g}
\phantomcaption\label{fig:fig3_h}
\phantomcaption\label{fig:fig3_i}
\phantomcaption\label{fig:fig3_j}
\end{subcaptiongroup}
\captionsetup{subrefformat=parens}
\caption{\label{fig:fig3} \justifying \textbf{Gate voltage, thermal cycling, and interface effects.} 
\subref*{fig:fig3_a} $\delta \sigma_{yxx}$ as a function of the out-of-plane magnetic field $H$, measured at various gate voltages $V_{\mathrm{G}}$. Consecutive curves are vertically offset by $10\ \mathrm{pmV^{-1}\Omega^{-1}}$ for clarity.
\subref*{fig:fig3_b} Cross-correlation between $\delta \sigma_{yxx}(H)$ measured at a reference gate voltage $V_{\mathrm{G}} = V_{\mathrm{G}}^{\mathrm{ref}}$ and $\delta \sigma_{yxx}(H)$ obtained at various $V_{\mathrm{G}}$ values, plotted as a function of $V_{\mathrm{G}}$. A characteristic correlation gate voltage of approximately $10$~V is observed. 
\subref*{fig:fig3_c} $\delta\sigma_{yxx}(H)$ measured at $T = 2.5$~K after consecutive thermal cycling up to a warming temperature $T_{\mathrm{w}}$. Consecutive curves are vertically offset by $10\ \mathrm{pmV^{-1}\Omega^{-1}}$ for clarity. The fluctuations exhibit progressively different features with increasing $T_{\mathrm{w}}$. 
\subref*{fig:fig3_d} Cross-correlation between $\delta\sigma_{yxx}(H)$ obtained at $T = 2.5$~K and $\delta\sigma_{yxx}(H)$ measured after thermal cycling up to $T_{\mathrm{w}}$, plotted as a function of $T_{\mathrm{w}}$. The fluctuations lose correlation at an onset temperature $T_{\mathrm{onset}} \simeq 40$~K. According to our proposed model, $T_{\mathrm{onset}}$ indicates the emergence of pinned ferroelectricity at the surface (see main text for discussion). The solid and dashed lines are guides to the eye. 
\subref*{fig:fig3_e} Schematic of the $\mathrm{KTaO}_3$(100)- and $\mathrm{KTaO}_3$(110)-based interfaces. 
\subref*{fig:fig3_f}--\subref*{fig:fig3_g} $\delta \sigma_{yxx}(H)$ measured at various temperatures for the $\mathrm{LaAlO}_3/\mathrm{KTaO}_3$(100) and $\mathrm{LaAlO}_3/\mathrm{KTaO}_3$(110) interfaces. Fluctuation patterns with similar magnitudes but distinct features are observed across all interface orientations. 
\subref*{fig:fig3_h} Comparison of the second-harmonic transverse voltage, $V_{yxx}^{2\omega}(H)$, for the $\mathrm{LaAlO}_3/\mathrm{KTaO}_3$(111) and $\mathrm{LaTiO}_3/\mathrm{SrTiO}_3$(111) heterostructures. 
\subref*{fig:fig3_i} Normalized $V_{yxx}^{2\omega}(H)$ curves, scaled by the signal amplitude at $H = 9$~T for clarity. 
\subref*{fig:fig3_j} $\Delta V_{yxx}^{2\omega}(H)$, extracted from the data marked by the dashed box in \subref*{fig:fig3_i}. To compare the interfaces, a linear trend was subtracted for simplicity. Solid and dashed lines represent up- and down-field sweeps, respectively. Reproducible fluctuations are conspicuous in $\mathrm{KTO}$, while they remain completely unobservable in $\mathrm{STO}$. Measurements for STO and KTO were taken at $V_{\mathrm{G}} = 4$~V ($T = 1.8$~K) and $V_{\mathrm{G}} = 24$~V ($T = 2.5$~K), respectively.} 
\end{figure*}

We next investigate the dependence of the fluctuation pattern on the magnetic field orientation, parameterized by the tilt angle $\theta$ (Fig. \ref{fig:fig2_a}). We observe that the fluctuations depend primarily on the out-of-plane magnetic field component, $H_\perp = H \cos\theta$ (Fig. \ref{fig:fig2_b}). This quasi-two-dimensional behavior is confirmed by tracking specific fluctuation features across different tilt angles; their positions in total field $H$ scale precisely with $1/\cos\theta$ (Fig. \ref{fig:fig2_c}). Interestingly, while the positions of the fluctuations are determined by $H_\perp$, their amplitude is sensitive to the in-plane component, $H_\parallel$. As $\theta$ increases, the fluctuation magnitude decays. This suppression at high tilt angles may arise from the relative shift of the Rashba-split Fermi surfaces induced by the parallel field \cite{tuvia2024enhanced}.
\par
To establish the electronic properties of the interface, we examine the linear longitudinal magnetoresistance (first harmonic) in Fig.~\ref{fig:fig2_d}. The data display characteristic weak anti-localization (WAL) behavior, typical of a 2D electron system with strong spin-orbit coupling \cite{bergmann1982weak} (see Supplementary Figs.~S6--S7). From these measurements, we extract the phase coherence length, $l_\phi$ (Fig.~\ref{fig:fig2_e}). Notably, the extracted phase coherence length ($l_\phi \simeq 100$~nm) is orders of magnitude smaller than the macroscopic device dimensions ($L=200\ \mathrm{\mu m}$). Furthermore, we observe a complete absence of fluctuations in the linear Hall voltage (see Supplementary Fig.~S4). This smoothness in the linear response indicates that standard linear universal conductance fluctuations (UCF) efficiently self-average to zero over the massive device area, ruling them out as the origin of the nonlinear patterns. Finally, while we cannot resolve gate-tuned fluctuations in the nonlinear longitudinal voltage, magnetic-field-tuned fluctuations remain detectable in this channel. We attribute these longitudinal fluctuations to the admixture of the transverse signal into the longitudinal measurement (see Supplementary Fig. S4).
\par

To determine the relevant length scale governing the surviving nonlinear fluctuations, we compute their magnetic field autocorrelation function (see Supplementary Fig. S5). By converting the extracted magnetic correlation field into a characteristic length scale ($l_f$) representing the effective coherent area, we find that $l_f$ is remarkably comparable to $l_\phi$ and shares its decreasing trend with temperature (Fig. \ref{fig:fig2_e}). This result strongly indicates that the nonlinear transport is not governed by the macroscopic geometrical dimensions of the device, but rather by a mesoscopic length scale.  
\par
Valuable insights into the origin of the nonlinear transport can be gleaned from its temperature dependence. As temperature increases, the variation in electron mobility dominates the longitudinal conductivity, $\sigma_{xx}$, while the Hall-density remains relatively constant (see Supplementary Fig. S13). At low temperature $T<60$ K, we find that the macroscopic nonlinear Hall response $\sigma_{yxx}$ scales predominantly with $\sigma_{xx}^3$, alongside a smaller contribution proportional to $\sigma_{xx}$ (Fig. \ref{fig:fig2_f}). In a stark contrast, the fluctuation strength, $\delta\sigma_{yxx}^{rms}$, does not exhibit a simple power-law scaling with $\sigma_{xx}$. Instead, the fluctuations drop precipitously with increasing temperature, becoming immeasurably small above $T \simeq 10$ K. This sharp thermal suppression, which decouples entirely from the more moderate temperature dependence of $\sigma_{xx}$ (Fig. S5), strongly argues against extrinsic scattering mechanisms as the primary driver of the fluctuation signal. Rather, this thermal suppression seems closely tied to the loss of quantum coherence, as $\delta\sigma_{yxx}^{rms}$ and $l_\phi$ are found to be proportional at low temperatures (Fig. \ref{fig:fig2_g}), a strong indication that the observed fluctuations arise from an interplay of quantum geometry and disorder.

\par
We investigate the crystalline anisotropy of the nonlinear response by varying the angle, $\psi$, between the driving current $I^\omega$ and the $[1\bar{1}0]$ reference axis of the (111) interface (see scheme in Fig. \ref{fig:fig2_h}). The macroscopic second-harmonic Hall voltage, $V_{yxx}^{2\omega}$, exhibits the dominant trigonal symmetry anticipated for the (111) surface (Fig. \ref{fig:fig2_i}), a behavior consistent with skew scattering as the leading bulk contribution \cite{he2021quantum, ZhaiNanoLett2023}. This symmetry is evident in the angular dependence: a maximal nonlinear voltage is measured when current flows along the $[1\bar{1}0]$ axis ($\psi =$ 0°), whereas the signal is significantly suppressed, as dictated by symmetry constraints, along the $[\bar{1}\bar{1}2]$ axis ($\psi =$ 90°). In striking contrast to this bulk behavior, the fluctuations in the nonlinear Hall conductivity ($\delta\sigma_{yxx}$) do not obey these symmetry restrictions (see Fig. \ref{fig:fig2_j}). They remain robust, exhibiting comparable magnitudes, albeit with distinct fluctuation patterns, at all angles, even along crystallographic directions where the macroscopic signal is strictly forbidden. This independence from the macroscopic crystal orientation suggests that the fluctuations are governed by the local disorder patterns rather than the long-range crystal symmetry.
\par
To explore the role of chemical potential tuning, we measure $\delta\sigma_{yxx}(H)$ across different gate voltages (Fig. \ref{fig:fig3_a}). The fluctuation patterns exhibit a high sensitivity to electrostatic gating. While we observe a strong correlation between traces measured at nearby gate voltages, the fluctuations gradually decorrelate when the gate voltage is varied over a scale of $\sim 10$ V (Fig. \ref{fig:fig3_b}). This rapid decorrelation points to the electronic origin of the fluctuations. We show below that this gate dependence arises as the chemical potential shifts through various massive Dirac nodes in momentum space; the resulting BCD is highly sensitive to the precise contour of the Fermi surface \cite{ma2019observation}. Furthermore, electrostatic tuning modifies the screening of the disorder potential itself, thereby altering the local Berry curvature landscape \cite{chen2024nonlinear}. In direct analogy to conventional UCF, this system is expected to exhibit a finite correlation energy scale \cite{lee1985universal}, a feature physically consistent with the finite gate-voltage correlation range observed in our measurements.
\par
To assess the role of the 2DES confinement direction, we measure the nonlinear Hall effect at the LAO/KTO (100) and (110) interfaces. In contrast to the (111) interface, which lacks in-plane inversion symmetry, these specific surface terminations preserve inversion symmetry, meaning that macroscopic second-order nonlinear transport should be strictly forbidden (illustrated in Fig. \ref{fig:fig3_e}). Nevertheless, we observe a finite macroscopic nonlinear Hall conductivity at these interfaces (see Supplementary Figs. S14-S15), in agreement with previous studies \cite{ZhaiNanoLett2023, krantz2024nonlinear}. More importantly, nonlinear Hall conductivity fluctuations, $\delta\sigma_{yxx}(H)$, are robustly observed across both of these orientations (Figs. \ref{fig:fig3_f} and \ref{fig:fig3_g}). These fluctuation patterns exhibit a magnitude comparable to those of the (111) interface—though their exact mesoscopic fingerprints differ, and display the same sharp thermal suppression, becoming immeasurably small above $T \simeq 10$~K. These observations point to a common physical origin for the fluctuations regardless of the macroscopic crystal face orientation, a behavior entirely consistent with local inversion symmetry breaking driven by pinned ferroelectric distortions.
\par
We interpret these observations as signatures of mesoscopic Berry curvature fluctuations arising from static disorder. We propose that this phenomenon is rendered conspicuous in KTO by the combination of strong $5d$ spin-orbit coupling and local inversion symmetry breaking, which is driven by a surface ferroelectric instability \cite{zhang2025enhanced}. Microscopically, oxygen vacancies residing in the conductive layer act as the primary disorder sites enabling this symmetry breaking \cite{dong2026strongly}. At low temperatures, these vacancies induce local structural distortions in the Ta atomic positions, thereby pinning the ferroelectric polarization. This local polarization restricts the point-group symmetry to a single mirror line (Fig. \ref{fig:summary_d}), a condition that strictly allows for the emergence of a Berry-curvature dipole (BCD) in momentum space oriented perpendicular to the mirror plane \cite{mercaldo2023orbital} (Fig. \ref{fig:summary_g}). Since the vacancy concentration is inherently linked to the 2DES electron density, which is roughly 10$^{13}$ cm$^{-2}$ \cite{LiuScience2021}, its distribution on the surface is dilute. Consequently, while the induced local electric dipoles may dynamically fluctuate at higher temperatures, they become statically pinned by the onset of the ferroelectric instability at low temperatures, creating a robust, disordered landscape of BCDs. A pictorial summary of this microscopic mechanism is presented in Fig. \ref{fig:summary_e}.
\par
To reveal the decisive role of ferroelectric distortions in shaping the observed nonlinear Hall fluctuations, we systematically investigate their dependence on thermal cycling. We first measure the fluctuation pattern, $\delta\sigma_{yxx}(H)$, at a base temperature of $2.5$~K (above the superconducting transition temperature $T_c=1.73\pm0.07~\mathrm{K}$, see Supplementary Fig. S10). We then perform a thermal cycle by warming the system to a target temperature, $T_\text{w}$, and cooling it back to $2.5$~K. This procedure is repeated for consecutively increasing values of $T_\text{w}$ up to room temperature (Fig. \ref{fig:fig3_c}). The cross-correlation between the initial $\delta\sigma_{yxx}(H)$ baseline and the pattern measured after each thermal cycle is shown in Fig. \ref{fig:fig3_d}. The fluctuations become uncorrelated at a characteristic temperature of $T_\text{onset} \simeq 40$~K, evidenced by a sharp drop in the cross-correlation amplitude. This $T_\text{onset}$ points to the thermal depinning of ferroelectric distortions, allowing the spatial arrangement of the local electric dipoles to reconfigure when $T > T_\text{onset}$, thereby reshaping the mesoscopic fluctuation landscape upon cooling. This energy scale is consistent with the reported onset of hysteresis in the $R(V_\text{G})$ curves, which reflects the emergence of ferroelectricity in the LAO/KTO (111) 2DES \cite{zhang2025enhanced}.
\par
To further elucidate the physical mechanism driving these mesoscopic fluctuations, it is instructive to compare the $\mathrm{KTO}$ surface 2DES with the well-studied 2DES hosted in $\mathrm{SrTiO_3}$ ($\mathrm{STO}$). Both of these perovskite oxides are quantum paraelectrics that host incipient ferroelectricity and exhibit a significant, gate-tunable spin-orbit interaction at their interfaces \cite{PhysRevLett.104.126802}. Figures \ref{fig:fig3_h} through \ref{fig:fig3_j} compare the second-harmonic Hall voltage, $V^{2\omega}_{yxx}$, measured in $\mathrm{LaAlO_3/KTaO_3(111)}$ and $\mathrm{LaTiO_3/SrTiO_3(111)}$. It is evident that while the macroscopic nonlinear signal is substantially larger in the $\mathrm{SrTiO_3}$-based device, possibly due to structural domains present in $\mathrm{SrTiO_3}$ \cite{persky2025imaging} absent in cubic $\mathrm{KTaO_3}$, the mesoscopic fluctuations are completely absent. The spin-orbit energy inferred from the weak anti-localization effect, $\Delta_\mathrm{so}\simeq 75$ meV (Supplementary Fig. S6), is roughly fifty times larger in $\mathrm{KTO}$, than the $\Delta_\mathrm{so}\simeq 1.5$ meV observed in $\mathrm{LaTiO_3/SrTiO_3(111)}$ \cite{tuvia2024enhanced}. This massive disparity arises from the strong atomic spin-orbit coupling of the heavy Ta $5d$ orbitals in $\mathrm{KTO}$ compared to the lighter Ti $3d$ orbitals in $\mathrm{SrTiO_3}$. These observations strongly suggest that the sheer magnitude of the spin-orbit interaction plays a critical role, rendering the nonlinear Hall conductivity fluctuations conspicuous in the $\mathrm{KTO}$ system while they remain unobservable in $\mathrm{SrTiO_3}$.
\par
The experimental observations indicate that the nonlinear conductivity fluctuations emerge in two steps. First, as the temperature drops below $T_{\text{onset}} \simeq$ 40 K, ferroelectric domains form, and Ta atoms adjacent to oxygen vacancies shift along spontaneously selected polarization directions. This breaking of in-plane mirror symmetry creates the conditions for a non-zero BCD to emerge within each domain. While the macroscopic expectation value of the BCD across a large enough device averages to zero, its local spatial variations drive the observed conductivity fluctuations. Because the BCD has the dimensions of length, the root-mean-square (RMS) of its spatial variations scales proportionally with a characteristic mesoscopic length. For the intrinsic nonlinear Hall signal, this length scale is dictated by the phase coherence length, $l_\phi$, in essence because the average number of inversion-breaking defect sites encompassed by a coherent electron wavefunction scales with $l_\phi^2$. Therefore, a second essential ingredient in the emergence of these nonlinear fluctuations is the sharp increase in $l_\phi$ below $\simeq$ 10 K. To analyze the effect of inversion-breaking disorder on the band structure, we treat a phase-coherent area of the crystal ($\sim l_\phi^2$) as a "supercell" within a periodic system, representing a single ferroelectric domain of area $\sim l_{FE}^2$. In the resulting mini-Brillouin zone, the original bands are heavily folded, generating numerous band crossings (see Fig. \ref{fig:summary_f}). The role of disorder is to open gaps at these crossings and break inversion symmetry, which manifests as a tilt in the dispersion near the avoided crossings (see Fig. \ref{fig:summary_g}). A conceptually similar approach to estimating disorder effects on the nonlinear Hall conductivity was utilized in Ref. \cite{chen2024nonlinear}. Another critical ingredient in this system is the strong spin-orbit coupling and the resulting band splitting, which separates band crossings with different Berry curvature in momentum and energy. The BCD is defined as
\begin{equation}
    D_i = \int \frac{d^2k}{(2\pi)^2} (\partial_{k_i} f_0(\mathbf{k})) \Omega(\mathbf{k})
\end{equation}
where $f_0(\mathbf k)$ is the Fermi-Dirac distribution and $\Omega(\mathbf{k})$ is the Berry curvature. Consequently, an asymmetric distribution of $\Omega$ on the Fermi surface results in a non-zero local BCD. This BCD, in turn, contributes to the second-order conductivity via $\delta\sigma_{yxx} = \frac{1}{2\hbar^2}\tau e^3D_x$ \cite{sodemann2015quantum}, where $\tau$ is the relaxation time. Denoting the RMS values of $D_x$ and $\delta\sigma_{yxx}$ as $D_x^{\text{rms}}$ and $\delta\sigma_{yxx}^{\text{rms}}$, respectively, we obtain $\delta\sigma_{yxx}^{\text{rms}} \approx \frac{l_{FE}}{\sqrt{LW}}\frac{\tau e^3}{2\hbar^2}D_x^{\text{rms}}$, where $W$ and $L$ are the width and length of the sample (see Supplementary Sec. I for further details). The magnitude of the BCD fluctuations, $D_x^{\text{rms}}$, can be estimated using a simplified band structure where each crossing is associated with a massive Dirac node tilted in the $k_x$ direction. This is described by the Hamiltonian $H = t v k_x \sigma_0 + v (k_x \sigma_x + k_y \sigma_y) \pm m \sigma_z $, where $v$ is the Fermi velocity, $m$ is the typical band gap, $t$ is the dimensionless tilt parameter, and $\sigma_0$ is the identity matrix. The resulting energy dispersion and Berry curvature distributions are plotted in Fig. \ref{fig:summary_g}. Assuming the node positions and the $\pm$ signs are distributed randomly and uniformly, and accounting for the band structure of the KTO (111) interface \cite{Bruno2019}, we predict (see Supplementary Sec. I) that $D^{\text{rms}}_x \approx 0.5 t^{\text{rms}} l_\phi$, where $t^{\text{rms}}$ is the RMS of the tilt parameter. Based on the data in Fig. \ref{fig:fig2}, we verify the proportionality $D_x^{\text{rms}} \propto l_\phi$ and estimate $l_{FE}t^{\text{rms}} \approx$ 1 $\mu$m. In addition to reproducing the proportionality $\delta \sigma_{yxx}^{\text{rms}} \propto l_\phi$, the $l_\phi^2$-supercell construction also explains the strong sensitivity of the BCD to the out-of-plane magnetic field. In a tight-binding Hamiltonian, each hopping term acquires a phase. Due to the large number of atoms per $l_\phi^2$ plaquette, even a relatively small magnetic field can drastically modify the band structure and, critically, the Berry curvature. Quantitatively, we observe that the characteristic length scale $l_f$ derived from the magnetic field dependence of $\delta\sigma_{yxx}$ exhibits the same order of magnitude and a similar temperature dependence as $l_\phi$ (Fig. \ref{fig:fig2_e}), fully consistent with our microscopic picture. Finally, while our calculations focus on the (111) termination, the band structures computed for the other surface terminations \cite{krantz2024nonlinear, ZhaiNanoLett2023, Bruno2019, Mallik2023electronic, Martinez2023, Zhang2019unusual, Varotto2022direct} also feature significant Rashba splitting and comparable Fermi velocities. This suggests that the identical physical mechanism is at play across all terminations, and the magnitude of the resulting effect should be comparable, which is indeed observed across all surface orientations we tested.
\par
We compare our findings with the gate-tunable nonlinear Hall fluctuations reported in corrugated bilayer graphene \cite{ho2021hall}. There, the fluctuating Berry curvature dipole was attributed solely to semi-classical shifts of the chemical potential through a statically disordered landscape of tilted mini-Dirac cones, and quantum interference was explicitly ruled out as the underlying mechanism. In stark contrast, our experiment reveals robust, magnetic-field-driven fluctuation patterns. This magnetic field dependence provides direct macroscopic evidence that the mesoscopic nonlinear response in our system fundamentally originates from quantum interference. Furthermore, whereas the graphene signal remains robust under thermal cycling, we observe a complete erasure of mesoscopic memory above $40\text{ K}$, establishing a direct link between this interference-driven nonlinear response and a dynamic, pinned ferroelectric landscape.
\par

In summary, we have demonstrated that second-order nonlinear transport in KTaO$_3$ interfaces exhibits conspicuous, reproducible mesoscopic fluctuations. Remarkably, these fluctuations persist even in macroscopically large devices. While the macroscopic nonlinear Hall effect adheres to the expected trigonal symmetry of the (111) surface, the mesoscopic fluctuations are ubiquitous across all studied crystal orientations, emerging robustly even along axes where the bulk nonlinear signal is strictly symmetry-forbidden. The extreme sensitivity of these patterns to electrostatic gating confirms their electronic origin, while their robust, magnetic-field-driven behavior provides direct macroscopic evidence of quantum interference. Quantitative analysis reveals that the characteristic length scale and the root-mean-square amplitude of these fluctuations scale directly with the phase coherence length ($l_\phi$), firmly establishing that the signal arises from quantum interference within independent, phase-coherent mesoscopic regions.
\par
We interpret these findings as the transport signature of a highly disordered, mesoscopic landscape of Berry curvature dipoles that is intimately tied to the interfacial lattice structure. In our proposed mechanism, this local structural symmetry breaking is driven by incipient ferroelectric polarization statically pinned by oxygen vacancies at the interface. This physical picture—linking quantum geometry directly to a structural phase—is strongly supported by the system's dual thermal response. Specifically, we observe a suppression of the fluctuation amplitude above $T \simeq 10$~K due to the loss of electronic phase coherence, alongside a complete erasure of the mesoscopic fluctuation memory upon thermal cycling above $T_\text{onset} \simeq 40$~K, a temperature that perfectly aligns with the onset of surface ferroelectricity in KTaO$_3$.
\par
More broadly, we demonstrate that quantum geometry can serve as an extraordinarily sensitive new probe for complex materials. By utilizing quantum interference in nonlinear transport—driven by the mesoscopic details of the Berry curvature—this geometric response allows for the diagnosis of hidden local structural orders, such as defect-pinned polar domains. Because these mesoscopic structural instabilities remain fundamentally invisible to linear transport and fall below the resolution of conventional bulk structural characterization, our approach offers a uniquely direct window into nanoscale symmetry breaking. By establishing a definitive link between local lattice instabilities and macroscopic nonlinear geometric transport, our work not only highlights KTaO$_3$ as a unique laboratory for tuning these effects, but establishes nonlinear mesoscopic transport as a powerful diagnostic tool for the design and characterization of next-generation quantum materials.
\section{Methods}
\subsection{Device fabrication}
Two-dimensional electron systems (2DESs) were realized at $\mathrm{LaAlO_3/KTaO_3}$ interfaces across various substrate orientations. First, the $\mathrm{KTaO_3}$ surfaces were treated with a buffered oxide etch (BOE) for $13$ min, followed by immersion in deionized water for $10$ min. To define the device geometry, the pre-etched substrates were patterned using maskless lithography (MLA). Non-conducting regions were defined by pulsed laser deposition (PLD) of an amorphous, $40$-nm-thick $\mathrm{BaTiO_3}$ layer, which served as a hard mask. To stabilize the surface, the samples were annealed at $600^\circ\mathrm{C}$ in an $\mathrm{O_2}$ partial pressure of $1 \times 10^{-4}$ Torr for $1.5$ h. To promote oxygen vacancies at the surface, the samples were subsequently annealed at $500^\circ\mathrm{C}$ and $1 \times 10^{-6}$ Torr for $30$ min. The active $6$-nm-thick $\mathrm{LaAlO_3}$ layer was then deposited via PLD at $500^\circ\mathrm{C}$ and $1 \times 10^{-6}$ Torr, using a laser fluence of $1.15\ \mathrm{J\,cm^{-2}}$ and a repetition rate of $2$ Hz. Electrical contacts to the buried 2DES were established using a manual wire bonder. The superconducting transitions for the surface orientations studied in this work are provided in Supplementary Fig. S10.
\subsection{Transport Measurements}
Magnetotransport measurements were performed in a cryogen-free Attocube attoDRY system equipped with a $9\text{ T}$ superconducting magnet. To investigate the angular dependence of the nonlinear response, the samples were mounted on a two-axis rotator, allowing for precise control of the magnetic field orientation relative to the sample normal ($\theta$) and the current direction ($\phi$). The nonlinear signals were measured using standard lock-in techniques. An alternating current $I_x(t) = \sqrt{2}I^{\omega}\sin(\omega t)$ was applied, where $I^\omega$ is the RMS amplitude of the current and $\omega=2\pi f$ is the frequency. The measurements were performed using $I^{\omega} = 50~\mathrm{\mu A}$ and $f = 11.189$~Hz. We verified that the measured response is independent of the excitation frequency in the low-frequency regime of $1$--$20$~Hz (Supplementary Fig.~S9), and that the fluctuation features are independent of the current amplitude (Supplementary Fig.~S8). The linear (first-harmonic) and nonlinear (second-harmonic) components of both the longitudinal and transverse voltages were acquired simultaneously using four separate lock-in amplifiers. The second-harmonic transverse voltage $V_{yxx}^{2\omega}$ was isolated to extract the nonlinear Hall conductivity $\sigma_{yxx}$ (see Supplementary Sec. II for further details), while the linear longitudinal voltage $V_{xx}^\omega$ was concurrently monitored to determine the phase coherence length $l_\phi$ via weak antilocalization analysis (Supplementary Sec. III.A).
\begin{acknowledgments} We thank Moshe Goldstein and Alon Ron for insightful discussions, and G. Tuvia for help with the STO sample. The experimental research is supported by the Israel Science Foundation under grant No. 1711/23. S. Sandik acknowledges support from the PBC quantum initiative. A portion of this work was performed at the National High Magnetic Field Laboratory, which is supported by the National Science Foundation Cooperative Agreement No. DMR-2128556 and the State of Florida. T.H. acknowledges financial support from the European Research Council (ERC) under grant QuantumCUSP (Grant Agreement No. 101077020). P.M. was supported by the Harry Bloomfield International Scholarship and a postdoctoral fellowship from the Azrieli Foundation.\end{acknowledgments}

\putbib[Ref]
\end{bibunit}

\clearpage
\newpage

\let\addcontentsline\savedaddcontentsline

\graphicspath{{figures_si/}}

\setcounter{page}{1}
\renewcommand{\thepage}{S\arabic{page}}
\setcounter{section}{0}
\setcounter{figure}{0}
\renewcommand{\thefigure}{S\arabic{figure}}
\setcounter{equation}{0}
\renewcommand{\theequation}{S\arabic{equation}}

\let\oldmaketitle\maketitle
\renewcommand{\maketitle}{
    \let\tempaddcontentsline\addcontentsline
    \renewcommand{\addcontentsline}[3]{} 
    \oldmaketitle                        
    \let\addcontentsline\tempaddcontentsline 
    \pdfbookmark[0]{Observation of Berry curvature fluctuations from incipient polar order in an oxide interface}{sibookmark}
}

\stepcounter{bibsectioncount}
\begin{bibunit}
    \title {Supplementary Information for: Observation of Berry curvature fluctuations from incipient polar order in an oxide interface}

\author{S. Avraham}
\affiliation{School of Physics and Astronomy, Tel Aviv University, Tel Aviv 6997801, Israel}

\author{D. Gitman}
\affiliation{School of Physics and Astronomy, Tel Aviv University, Tel Aviv 6997801, Israel}

\author{P. Matus}
\affiliation{School of Physics and Astronomy, Tel Aviv University, Tel Aviv 6997801, Israel}

\author{S. Sandik}
\affiliation{School of Physics and Astronomy, Tel Aviv University, Tel Aviv 6997801, Israel}

\author{E. Raz}
\affiliation{School of Physics and Astronomy, Tel Aviv University, Tel Aviv 6997801, Israel}

\author{S. Jana}
\affiliation{School of Physics and Astronomy, Tel Aviv University, Tel Aviv 6997801, Israel}

\author{M. Dahan}
\affiliation{School of Physics and Astronomy, Tel Aviv University, Tel Aviv 6997801, Israel}

\author{T. Holder}
\email{tobiasholder@tauex.tau.ac.il}
\affiliation{School of Physics and Astronomy, Tel Aviv University, Tel Aviv 6997801, Israel}

\author{Y. Dagan}
\email{yodagan@tauex.tau.ac.il}
\affiliation{School of Physics and Astronomy, Tel Aviv University, Tel Aviv 6997801, Israel}

\date{\today}

\maketitle

\tableofcontents       
\listoffigures
\clearpage

\section{Theoretical model}

\begin{figure}[h]
    \includegraphics[width=0.6\textwidth]{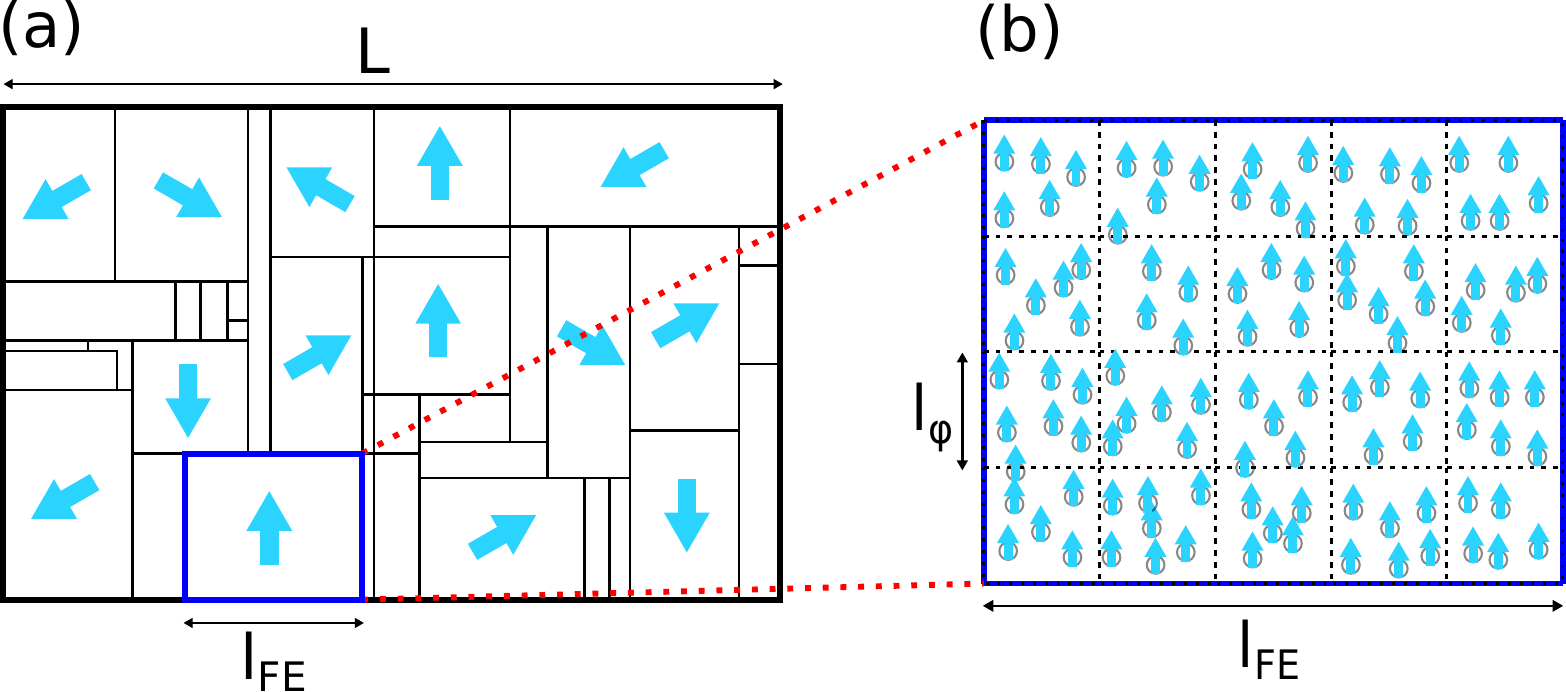}
    \caption[Hierarchy of length scales]{\label{fig:domains} \justifying \textbf{Hierarchy of length scales}. (a) The system of size $L$ is divided into ferroelectric-like domains. The blue arrows show the direction of polarization, and the length scale $l_{FE}$ represents the typical size of the larger domains, since they contribute the most to the total conductivity. (b) These domains, in turn, can be divided into many smaller regions of size $l_\phi$, where electrons maintain coherence. Finally, each coherent region contains many oxygen vacancies, symbolically denoted by circles, which act as centers of ferroelectric distortions shown as blue arrows. 
    }
\end{figure}

The fluctuations in the nonlinear Hall effect are observable for temperatures $T\lessapprox 10$~K. As discussed in the main text, within this temperature range, the following hierarchy can be established between the different length scales in the system (see Fig.~\ref{fig:domains}):\begin{equation}a \ll l_{\mathrm{dis}} \ll l_\phi \ll l_{\mathrm{FE}} < L,\end{equation}where $a$ is the lattice constant, $l_{\mathrm{dis}}$ is the average distance between oxygen vacancies, $l_\phi$ is the phase coherence length, $l_{\mathrm{FE}}$ is the correlation length of the ferroelectric distortions, and $L$ is the sample size. The intrinsic contribution to the nonlinear Hall conductivity $\delta\sigma_{yxx}$ is typically related to the Berry curvature dipole (BCD) $D_i$ in the following way:
\begin{equation}
    \delta\sigma_{yxx} = \frac{e^3 \tau}{2\hbar^2} D_x, \qquad D_i = \int \frac{d^2k}{(2\pi)^2} f_0(\mathbf{k}) \frac{\partial \Omega(\mathbf{k})}{\partial k_i},
\end{equation}
where $\tau$ is the momentum relaxation time, $f_0(\mathbf{k})$ is the Fermi-Dirac distribution, and $\Omega(\mathbf{k})$ is the Berry curvature \cite{sodemann2015quantum}. A complicating factor in the present case is that the inversion-breaking distortions are centered on randomly located oxygen vacancy sites. This means that to properly evaluate the effect of symmetry breaking, one must account for the disordered, amorphous spatial landscape of these defects (see Fig.~\ref{fig:domains}). However, even though such a landscape is not strictly periodic, and thus globally does not possess a well-defined band structure or Berry curvature, we propose the following approach to estimate the effective BCD. Since a single electron wavefunction maintains its phase coherence only over a characteristic distance $l_\phi$, it is physically justified to divide a single macroscopic ferroelectric domain of size $\sim l_{\mathrm{FE}}^2$ into an ensemble of smaller subsystems of size $\sim l_\phi^2$, within which quantum interference effects remain robust. In the following, we focus on analyzing the properties of these phase-coherent subsystems. To this end, we treat a single area of size $\sim l_\phi^2$ as a ``supercell'' of an artificial periodic system, which allows us to formulate well-defined statements regarding its effective band structure and local Berry curvature distribution.

\begin{figure}
    \begin{subcaptiongroup}
    \includegraphics[width=0.6\textwidth]{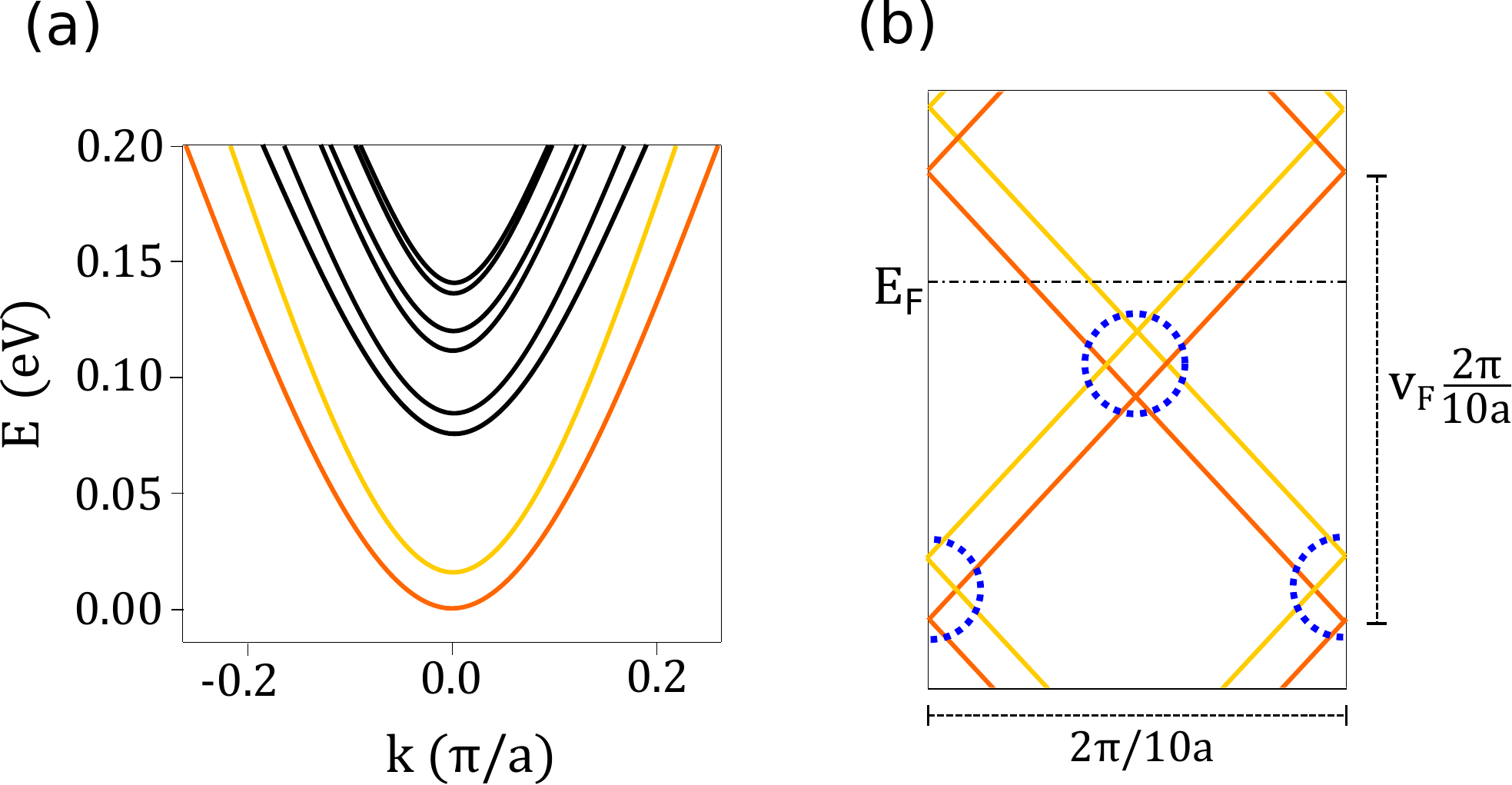}
    \phantomcaption\label{fig:figs0_a}
    \phantomcaption\label{fig:figs0_b}
    \end{subcaptiongroup}
    \captionsetup{subrefformat=parens}
    \caption[Band structure of $\mathrm{KTaO_3(111)}$ 2D electron system]{\label{fig:figs0} \justifying \textbf{Band structure of $\mathbf{KTaO_3(111)}$ 2D electron system}. \subref*{fig:figs0_a} Schematic energy-momentum dispersion along the $\Gamma-K$ direction, based on ~\cite{Bruno2019}. We focus on the outermost pair of bands, colored in orange and yellow. \subref*{fig:figs0_b} Energy-momentum dispersion of a supercell consisting of 10$\times$10 original unit cells along the $\Gamma-K$ direction. For clarity, only the folded bands that originate in the outermost pair of bands in Fig.~\ref{fig:figs0_a} are shown. The blue dashed circles show the locations of band crossings.
    }
\end{figure}

To better understand the behavior of this system, we utilize the band structure for the (111) termination reported in Ref.~\cite{Bruno2019}. As schematically illustrated in Fig.~\ref{fig:figs0_a} based on the results of Ref.~\cite{Bruno2019}, the spin-orbit coupling induces a strong Rashba splitting of the outermost band, with a splitting energy $\Delta_{so} \gg 10$~meV. This splitting is most pronounced along the $\Gamma-K$ path of the Brillouin zone, as well as along its two symmetry-equivalent directions. The maximum Fermi momentum in Fig.~\ref{fig:figs0_a} spans approximately one-fifth of the Brillouin zone. Consequently, as shown in Fig.~\ref{fig:figs0_b}, by constructing a supercell of $10\times10=100$ unit cells, we identify 8 band crossings within an energy window of $\sim v_F\frac{2\pi}{10a}$ (where $v_F$ is the Fermi velocity). This number is multiplied by a factor of 3 to account for the symmetry-equivalent $\Gamma-K$ directions. 

Scaling the supercell area by a factor of $(l_\phi/10a)^2$ yields approximately $(l_\phi/10a)^4$ times more crossings within the same energy interval. Concurrently, the area of the corresponding folded Brillouin zone decreases by a factor of $(l_\phi/10a)^2$. Thus, we estimate the density of band crossings per unit energy as:
\begin{equation}
    \frac{(\mathrm{band~crossings})}{(\mathrm{energy})} \sim 3\times8\times\frac{10a}{2\pi v_F}\times\left(\frac{l_\phi}{10a}\right)^{2}\approx 0.4\frac{l_\phi^2}{a v_F}.
\end{equation}

Disorder opens gaps at these crossing points. We model each avoided crossing as a massive Dirac node tilted along the $k_x$ direction:
\begin{equation}
    H = t v_F k_x + v_F \left( k_x \sigma_x + k_y \sigma_y \right) \pm \frac{1}{2}m\sigma_z,
    \label{eq:hamilt}
\end{equation}
where $m$ is the gap size and $t$ is the dimensionless tilt parameter. In the limit of small $t$, we find the Berry curvature dipole (BCD) contribution from a single avoided crossing as:
\begin{equation}
    D_x(E_F) \approx \pm\frac{3mtv_F(E_F^2-m^2)}{8\pi E_F^4},\qquad\mathrm{if}~~|E_F|\geq m.
\end{equation}
Here, $E_F$ is the energy difference between the Fermi level and the center of the gap. Assuming that the contributions from different band crossings to the overall BCD are additive and uncorrelated, we calculate the variance of $D_x$ as:
\begin{equation}
    \mathrm{Var}(D_x) = \frac{(\mathrm{band~crossings})}{(\mathrm{energy})} \times \int\! dE_F D_x(E_F)^2 P_t(t) dt = (t^{\mathrm{rms}})^2 \times 0.4 \times \frac{3}{70 \pi^2} \frac{l_\phi^2}{a v_F} \frac{v_F^2}{m},
    \label{eq:variance}
\end{equation}
where $P_t(t)$ is the probability distribution of the tilt $t$, and $t^{\mathrm{rms}}$ is the root-mean-square magnitude of the tilt. In the above equations, $v_F$ denotes the band slope $\partial E/\partial k$ (i.e. $\hbar$ times the physical Fermi velocity). Taking $E_F \approx 0.15$~eV \cite{Bruno2019}, we extract a Fermi velocity of $v_F \approx 400$~meV$\cdot a$ from Fig.~\ref{fig:figs0_a}. Using a gap size of $m \approx 3$~meV (consistent with the $40$ K temperature scale observed in Fig. 3d of the main text), we obtain the standard deviation:
\begin{equation}
    D_x^{\mathrm{rms}} \approx 0.5 t^{\mathrm{rms}} l_\phi.
    \label{eq:Drms}
\end{equation}

At this point, two remarks are in order. First, the assumption that the Hamiltonian parameters in Eq.~(\ref{eq:hamilt}) are uncorrelated across different band crossings is not strictly justified. In particular, in the limit of very weak spin-orbit coupling ($\Delta_{so} \ll m$), the quadruplets of crossings in Fig.~\ref{fig:figs0_b} converge to a single point with a vanishing total Berry curvature. As the ratio $\Delta_{so}/m$ increases, the BCD contributions from nearby crossings must remain partially anticorrelated. For this reason, the result in Eq.~(\ref{eq:Drms}) is likely an overestimation. Second, while we have focused on the (111) interface, the band structures for the other interface orientations---as computed in Refs.~\cite{krantz2024nonlinear, ZhaiNanoLett2023,Bruno2019,Mallik2023electronic,Martinez2023,Zhang2019unusual,Varotto2022direct}---also feature at least one pair of bands with a significant Rashba splitting and comparable Fermi velocities. Thus, while the exact numerical coefficient in Eq.~(\ref{eq:Drms}) may vary for other crystal orientations, the order of magnitude remains robust.

\begin{figure}
    \includegraphics[width=0.5\textwidth]{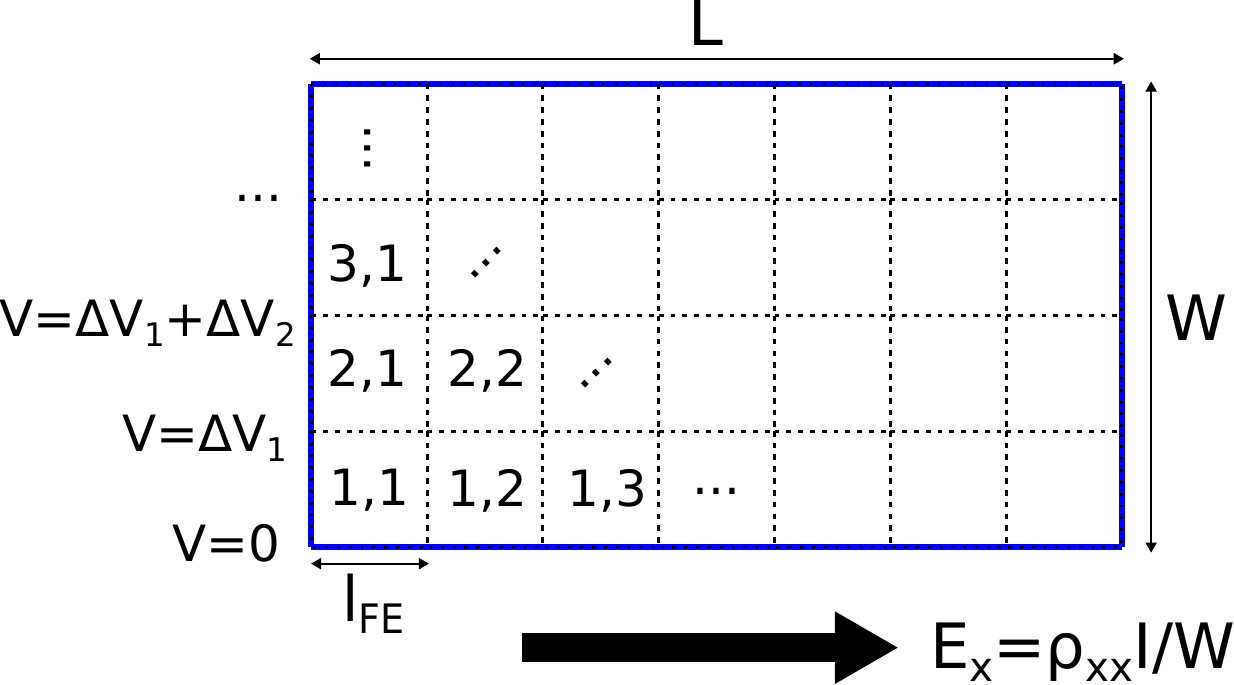}
    \caption[Calculating the transverse voltage]{\label{fig:domains2} \justifying \textbf{Calculating the transverse voltage}. The sample of length $L$ and width $W$ is divided into domains of size $l_{FE}^2$. The individual domains are labeled by a pair of indices $i,j$ as in the figure. The longitudinal electric field $E_x$ is proportional to the current $I$ through the sample. The method of calculating contributions to the transverse voltage $\Delta V_i$ is explained in the text.
    }
\end{figure}

As discussed earlier, the entire sample of length $L$ and width $W$ can be divided into ferroelectric-like domains of characteristic area $\sim l_{\mathrm{FE}}^2$. We label these domains with indices $i,j$ as in Fig.~\ref{fig:domains2}. Because the disorder behaves similarly across all regions of size $\sim l_\phi^2$ within a single domain (in particular, the direction of the ferroelectric distortions remains uniform), we expect that the local second-order conductivity $\sigma_{yxx,ij}$ will also be roughly constant throughout the entire domain area of $\sim l_{\mathrm{FE}}^2$. 

We estimate the resulting transverse voltage drop across the sample as follows. First, we divide the sample into parallel stripes of length $L$ and width $l_{\mathrm{FE}}$, and estimate the transverse voltage drop $\Delta V_i$ across a single stripe. Because the net transverse current must be zero in the steady state, we have:
\begin{equation}
    \Delta V_i\left(\frac{\rho_{xx}l_{\mathrm{FE}}} {L}\right)^{-1} = -E_x^2 l_{\mathrm{FE}} \sum_{j}\sigma_{yxx,ij}.
\end{equation}
Since the local conductivities $\sigma_{yxx,ij}$ are uncorrelated random variables, their variances are additive. Summing over the $L/l_{\mathrm{FE}}$ domains along the length of the stripe yields a standard deviation of:
\begin{equation}
    \Delta V_i^{\mathrm{rms}} \approx \frac{E_x^2}{\rho_{xx}}\frac{l_{\mathrm{FE}}^2}{L}\sqrt{\frac{L}{l_{\mathrm{FE}}}}\frac{\tau e^3}{2\hbar^2}D_x^{\mathrm{rms}}.
\end{equation}
To find the total transverse voltage drop across the entire sample, we must sum the independent random variables $\Delta V_i$ across the $W/l_{\mathrm{FE}}$ stripes connected in series along the transverse direction. Consequently, the standard deviation of the total voltage drop is:
\begin{equation}
    \Delta V^{\mathrm{rms}} \approx \frac{E_x^2 W}{\rho_{xx}}\frac{l_{\mathrm{FE}}}{\sqrt{LW}}\frac{\tau e^3}{2\hbar^2}D_x^{\mathrm{rms}}.
    \label{eq:DeltaV}
\end{equation}
This can be compared with the expected macroscopic voltage arising from a spatially uniform nonlinear conductivity background, $\bar{\sigma}_{yxx}$:
\begin{equation}
    \bar{V} = \frac{E_x^2 W}{\rho_{xx}}\bar{\sigma}_{yxx}. 
    \label{eq:V}
\end{equation}
Comparing Eqs.~(\ref{eq:DeltaV}) and~(\ref{eq:V}) reveals that the magnitude of the measured macroscopic fluctuations in the effective nonlinear conductivity, $\delta\sigma_{yxx}^{\mathrm{rms}}$, is given by:
\begin{equation}
    \delta\sigma_{yxx}^{\mathrm{rms}} = \frac{l_{\mathrm{FE}}}{\sqrt{LW}}\frac{\tau e^3}{2\hbar^2}D_x^{\mathrm{rms}}.
\end{equation}
To extract $D_x^{\mathrm{rms}}$ from the experimental data, we make use of the Drude formula for linear conductivity:
\begin{equation}
    \sigma_{xx} = \tau e^2\sum_{i\in\mathrm{occ.}}\frac{n_i}{m^*_i},
\end{equation}
where $n_i$ is the electron density of band $i$, $m^*_i$ is its effective mass, and the sum runs over all occupied conduction bands. The density $n_i$ for a single 2D band is given by $n_i = E_{F,i}\frac{m^*_i}{\pi\hbar^2}$. Combining these formulas, we obtain the following relation:
\begin{equation}
    \frac{l_{\mathrm{FE}}}{\sqrt{LW}}D_x^{\mathrm{rms}} = \frac{2}{\pi}\frac{\delta\sigma^{\mathrm{rms}}_{yxx}}{\sigma_{xx}}\frac{\sum_{i\in\mathrm{occ.}} E_{F,i}}{e}. 
    \label{eq:D_formula}
\end{equation}
Based on Refs.~\cite{Bruno2019,ZhaiNanoLett2023}, we estimate $\sum_{i\in\mathrm{occ.}} E_{F,i} \approx 0.5$~eV. From the experimental data shown in Figs.~2f-g at $T=2.5$~K, we have:
\begin{equation}
    \sigma_{xx} \approx 0.95~\mathrm{mS},\qquad \delta\sigma^{\mathrm{rms}}_{yxx} \approx 2.2~\mathrm{pmV^{-1}\Omega^{-1}}, \qquad l_\phi \approx 130~\mathrm{nm}.
\end{equation}
For our device dimensions ($L=200~\mu\mathrm{m}$, $W=96~\mu\mathrm{m}$), and substituting the expression for $D_x^{\mathrm{rms}}$ from Eq.~(\ref{eq:Drms}), we confirm the proportionality $\delta\sigma_{yxx}^{\mathrm{rms}} \propto l_\phi$ and extract the characteristic parameter factor:
\begin{equation}
    l_{\mathrm{FE}} t^{\mathrm{rms}} \approx 1.5~\mu\mathrm{m}.
\end{equation}

\section{Nonlinear Hall fluctuations}
\subsection{Fluctuations under time-reversal symmetry}
To extract the nonlinear conductivity, we solve the two fundamental equations describing the linear and nonlinear transport: 
\begin{equation}
    E_i = \rho_{ij}J_j + \rho_{ijk}J_j J_k,
    \label{eq:E}
\end{equation}
\begin{equation}
    J_i = \sigma_{ij}E_j + \sigma_{ijk}E_j E_k.
    \label{eq:J}
\end{equation}
The relevant part of the nonlinear conductivity tensor is:
\begin{equation}
    \sigma_{ijk}= \begin{bmatrix}
  \sigma_{xxx} & \sigma_{xxy} & \sigma_{xyx} & \sigma_{xyy}\\
  \sigma_{yxx} & \sigma_{yxy} & \sigma_{yyx} & \sigma_{yyy}
\end{bmatrix}.
\end{equation}
Substituting Eq.~(\ref{eq:E}) into Eq.~(\ref{eq:J}) yields:
\begin{equation}
    J_i = \sigma_{ia}(\rho_{ab}J_b + \rho_{abc}J_b J_c) + \sigma_{ide}(\rho_{df}J_f + \rho_{dfg}J_f J_g)(\rho_{eh}J_h + \rho_{ehl}J_h J_l).
\end{equation}
Equating terms of the same order in $J_i$, we find for the first order:
\begin{equation}
    \sigma_{ia}\rho_{ab} = \delta_{ib},
    \label{eq:linear}
\end{equation}
and for the second order:
\begin{equation}
    \sigma_{ia}\rho_{abc} + \sigma_{ide}\rho_{db}\rho_{ec} = 0.
\end{equation}
Multiplying this second-order equation by $\sigma_{ck}\sigma_{bj}$ and using the identity from Eq.~(\ref{eq:linear}), we obtain:
\begin{equation}
    \sigma_{ia}\rho_{abc}\sigma_{ck}\sigma_{bj} + \sigma_{ijk} = 0,
\end{equation}
\begin{equation}
    \sigma_{ijk} = -\sum_{a,b,c}\sigma_{ia}\rho_{abc}\sigma_{bj}\sigma_{ck}.
    \label{eq:nonlinear}
\end{equation}
In our experimental geometry, we drive a current along the $x$-direction and measure the resulting voltages in the $x$- and $y$-directions. Therefore, the relevant electric fields are:
\begin{equation}
    E_x^{(2)} = \rho_{xxx}J_x^2,
    \label{eq:rhoxxx}
\end{equation}
\begin{equation}
    E_y^{(2)} = \rho_{yxx}J_x^2,
    \label{eq:rhoyxx}
\end{equation}
where $E_i^{(1)}$ and $E_i^{(2)}$ are the linear and nonlinear electric fields, respectively, which satisfy $E_i = E_i^{(1)} + E_i^{(2)}$ for $i \in \{x,y\}$.

Imposing three-fold symmetry ($C_{3v}$) on the resistivity tensor, the non-zero elements are constrained as follows:
\begin{equation}
    \rho_{ijk}=\begin{bmatrix}
  \rho_{xxx} & \rho_{yxx} & \rho_{yxx} & -\rho_{xxx}\\
  \rho_{yxx} & -\rho_{xxx} & -\rho_{xxx} & -\rho_{yxx}
\end{bmatrix}.
\end{equation}
Having defined all elements of the nonlinear resistivity tensor, we calculate the specific nonlinear conductivity components using Eq.~(\ref{eq:nonlinear}) alongside the linear relations $\sigma_{yx} = -\sigma_{xy}$ and $\sigma_{xx} = \sigma_{yy}$:
\begin{equation}
    \sigma_{xxx} = -(\rho_{xxx}\sigma_{xx} - \rho_{yxx}\sigma_{xy})(\sigma_{xx}^2 + \sigma_{xy}^2),
    \label{eq:sigmaxxx}
\end{equation}
\begin{equation}
    \sigma_{yxx} = -(\rho_{xxx}\sigma_{xy} + \rho_{yxx}\sigma_{xx})(\sigma_{xx}^2 + \sigma_{xy}^2).
    \label{eq:sigmayxx}
\end{equation}
In the limit of time-reversal symmetry where $\sigma_{xy} = 0$, these expressions simplify significantly to:
\begin{equation}
    \sigma_{xxx} = -\rho_{xxx}\sigma_{xx}^3 \approx -\frac{V_{xxx}^{(2)}}{(V_{xx}^{(1)})^3} \frac{L^2}{W} I_x,
\end{equation}
\begin{equation}
    \sigma_{yxx} = -\rho_{yxx}\sigma_{xx}^3 \approx -\frac{V_{yxx}^{(2)}}{(V_{xx}^{(1)})^3} \frac{L^3}{W^2} I_x.
\end{equation}
Here, we have substituted Eqs.~(\ref{eq:rhoxxx}) and (\ref{eq:rhoyxx}), utilizing the experimental definitions $E_x^{(2)} = V_{xxx}^{(2)}/L$, $E_y^{(2)} = V_{yxx}^{(2)}/W$, and $J_x = I_x/W$, alongside the linear transport approximations $\sigma_{xx} \approx (L/W)(I_x/V_{xx}^{(1)})$ and $\sigma_{yx} \approx (V_{yx}^{(1)}/(V_{xx}^{(1)})^2)(L^2/W^2)I_x$. In these relations, $V_{xx}^{(1)}$ and $V_{xxx}^{(2)}$ denote the measured linear and nonlinear longitudinal voltages, respectively, while $V_{yx}^{(1)}$ and $V_{yxx}^{(2)}$ denote the measured linear and nonlinear transverse voltages. $I_x$ is the applied current, $W$ is the width, and $L$ is the length of the Hall bar device. 

Finally, the nonlinear Hall conductivity fluctuations are given by:
\begin{equation}
\label{eq:fluctEq}
    \delta\sigma_{yxx} = -\frac{L^3}{W^2}\frac{I_x \delta V_{yxx}^{(2)}}{(V_{xx}^{(1)})^3}.
\end{equation}

\subsection{Fluctuations under broken time-reversal symmetry}
In the presence of an external magnetic field (broken time-reversal symmetry), we obtain the components of the nonlinear conductivity tensor by substituting the linear approximations from above into Eq.~(\ref{eq:sigmaxxx}) and Eq.~(\ref{eq:sigmayxx}). In the small Hall angle limit ($V_{yx}^{(1)} \ll V_{xx}^{(1)}$), we can drop the higher-order terms, yielding:
\begin{equation}
    \sigma_{xxx} \approx -\frac{V_{xxx}^{(2)}}{(V_{xx}^{(1)})^3} \frac{L^2}{W} I_x \left[1 + \left(\frac{L}{W}\frac{V_{yx}^{(1)}}{V_{xx}^{(1)}}\right)^2\right]- V_{yxx}^{(2)}\frac{V_{yx}^{(1)}}{(V_{xx}^{(1)})^4} \frac{L^4}{W^3} I_x \left[1 + \left(\frac{L}{W}\frac{V_{yx}^{(1)}}{V_{xx}^{(1)}}\right)^2\right] \approx -\frac{V_{xxx}^{(2)}}{(V_{xx}^{(1)})^3} \frac{L^2}{W} I_x,
\end{equation}
\begin{equation}
\begin{split}
    \sigma_{yxx} &\approx V_{xxx}^{(2)}\frac{V_{yx}^{(1)}}{(V_{xx}^{(1)})^4} \frac{L^3}{W^2}I_x\left[1 + \left(\frac{L}{W}\frac{V_{yx}^{(1)}}{V_{xx}^{(1)}}\right)^2\right] - \frac{V_{yxx}^{(2)}}{(V_{xx}^{(1)})^3} \frac{L^3}{W^2}I_x\left[1 + \left(\frac{L}{W}\frac{V_{yx}^{(1)}}{V_{xx}^{(1)}}\right)^2\right] \\
    &\approx V_{xxx}^{(2)}\frac{V_{yx}^{(1)}}{(V_{xx}^{(1)})^4} \frac{L^3}{W^2}I_x - \frac{V_{yxx}^{(2)}}{(V_{xx}^{(1)})^3}\frac{L^3}{W^2}I_x \\
    &= \frac{L^3}{W^2}\frac{I_x}{(V_{xx}^{(1)})^3} \left( V_{xxx}^{(2)}\frac{V_{yx}^{(1)}}{V_{xx}^{(1)}} - V_{yxx}^{(2)} \right).
\end{split}
\end{equation}

We find two terms contributing to $\sigma_{yxx}$. However, experimental analysis reveals that the dominant contribution to the mesoscopic fluctuations arises almost entirely from the second term, $-\frac{V_{yxx}^{(2)}}{(V_{xx}^{(1)})^3}\frac{L^3}{W^2}I_x$. Therefore, we can safely continue to use Eq.~(\ref{eq:fluctEq}) to extract the fluctuation signal even in the presence of a magnetic field. 

Finally, the overall sign of the nonlinear conductivity depends on the specific spatial direction of the inversion-symmetry breaking with respect to the crystal axes. For convenience and clarity, throughout this work, we present the data without the global minus sign.

\subsection{Extracting fluctuations from measurements}
In this experiment, magnetotransport measurements are performed on Hall bar devices using a standard 6-point measurement configuration. The length and width of the active device channel are $L=200\ \mathrm{\mu m}$ and $W=96\ \mathrm{\mu m}$, respectively. We apply a low-frequency ac drive current and simultaneously measure the longitudinal and transverse voltage responses. Each voltage signal is split into two lock-in amplifiers: one configured to detect the first-harmonic voltage, and the other configured to detect the second-harmonic voltage.

The applied ac current is given by:
\begin{equation}
    I_x(t) = I^{\omega,\mathrm{pk}} \sin(\omega t).
    \label{eq:current}
\end{equation}
Considering the transverse voltage measurement, the system exhibits both linear and nonlinear responses:
\begin{equation}
    V_{yx}(t) = R_{yx}I_x(t) + R_{yxx} [I_x(t)]^2 = V_{yx}^{(1)}(t) + V_{yxx}^{(2)}(t).
\end{equation}
Substituting the ac current from Eq.~(\ref{eq:current}) yields:
\begin{equation}
\begin{split}
    V_{yx}(t) &= R_{yx}I^{\omega,\mathrm{pk}} \sin(\omega t) + R_{yxx} (I^{\omega,\mathrm{pk}})^2 \sin^2(\omega t) \\
    &= R_{yx}I^{\omega,\mathrm{pk}} \sin(\omega t) + \frac{1}{2} R_{yxx} (I^{\omega,\mathrm{pk}})^2 [1-\cos(2\omega t)] \\
    &= \frac{1}{2} R_{yxx} (I^{\omega,\mathrm{pk}})^2 + R_{yx}I^{\omega,\mathrm{pk}} \sin(\omega t) + \frac{1}{2} R_{yxx} (I^{\omega,\mathrm{pk}})^2 \sin\left(2\omega t - \frac{\pi}{2}\right).
\end{split}
\end{equation}
From this expansion, the peak amplitude of the first-harmonic voltage detected by the lock-in amplifier is:
\begin{equation}
    V_{yx}^{\omega,\mathrm{pk}} = R_{yx}I^{\omega,\mathrm{pk}}.
\end{equation}
Similarly, the peak amplitude of the measured second-harmonic voltage is:
\begin{equation}
    V_{yxx}^{2\omega,\mathrm{pk}} = \frac{1}{2} R_{yxx} (I^{\omega,\mathrm{pk}})^2.
    \label{eq:Vyxx_mes}
\end{equation}
Notice that the measured second-harmonic voltage amplitude corresponds to exactly half of the intrinsic nonlinear voltage term. Thus, the net nonlinear voltage coefficient is $V_{yxx}^{(2)} = R_{yxx} (I^{\omega,\mathrm{pk}})^2 = 2V_{yxx}^{2\omega, \mathrm{pk}}$. 

To calculate the nonlinear conductivity using data directly output by the lock-in amplifiers, it is necessary to convert the peak values ($I^{\omega,\mathrm{pk}}$, $V_{ij}^{\omega,\mathrm{pk}}$, and $V_{ijk}^{2\omega,\mathrm{pk}}$) into root-mean-square (rms) values:
\begin{equation}
    I^{\omega,\mathrm{pk}} = \sqrt{2} I^{\omega},
    \label{eq:I}
\end{equation}
\begin{equation}
    V_{ij}^{\omega, \mathrm{pk}} = \sqrt{2} V_{ij}^{\omega},
    \label{eq:Vij}
\end{equation}
\begin{equation}
    V_{ijk}^{(2)} = 2V_{ijk}^{2\omega, \mathrm{pk}} = 2\sqrt{2} V_{ijk}^{2\omega}.
    \label{eq:Vijk}
\end{equation}
Substituting Eqs.~(\ref{eq:I})--(\ref{eq:Vijk}) into Eq.~(\ref{eq:fluctEq}), we arrive at the final working formula:
\begin{equation}
    \delta\sigma_{yxx} = -\sqrt{2} \frac{L^3}{W^2} \frac{I^\omega \delta V_{yxx}^{2\omega}}{(V_{xx}^{\omega})^3},
    \label{eq:flucteqfixed}
\end{equation}
where $I^{\omega}$ and $V_{xx}^{\omega}$ are the rms amplitudes of the first-harmonic applied current and longitudinal voltage, respectively, and $\delta V_{yxx}^{2\omega}$ represents the rms amplitude of the second-harmonic Hall voltage fluctuations measured by the lock-in amplifier. As mentioned earlier, throughout this work, we present the data without the global minus sign.

\subsection{Fitting procedure}
\begin{figure*}
\begin{subcaptiongroup}
\includegraphics[width=.8\columnwidth]{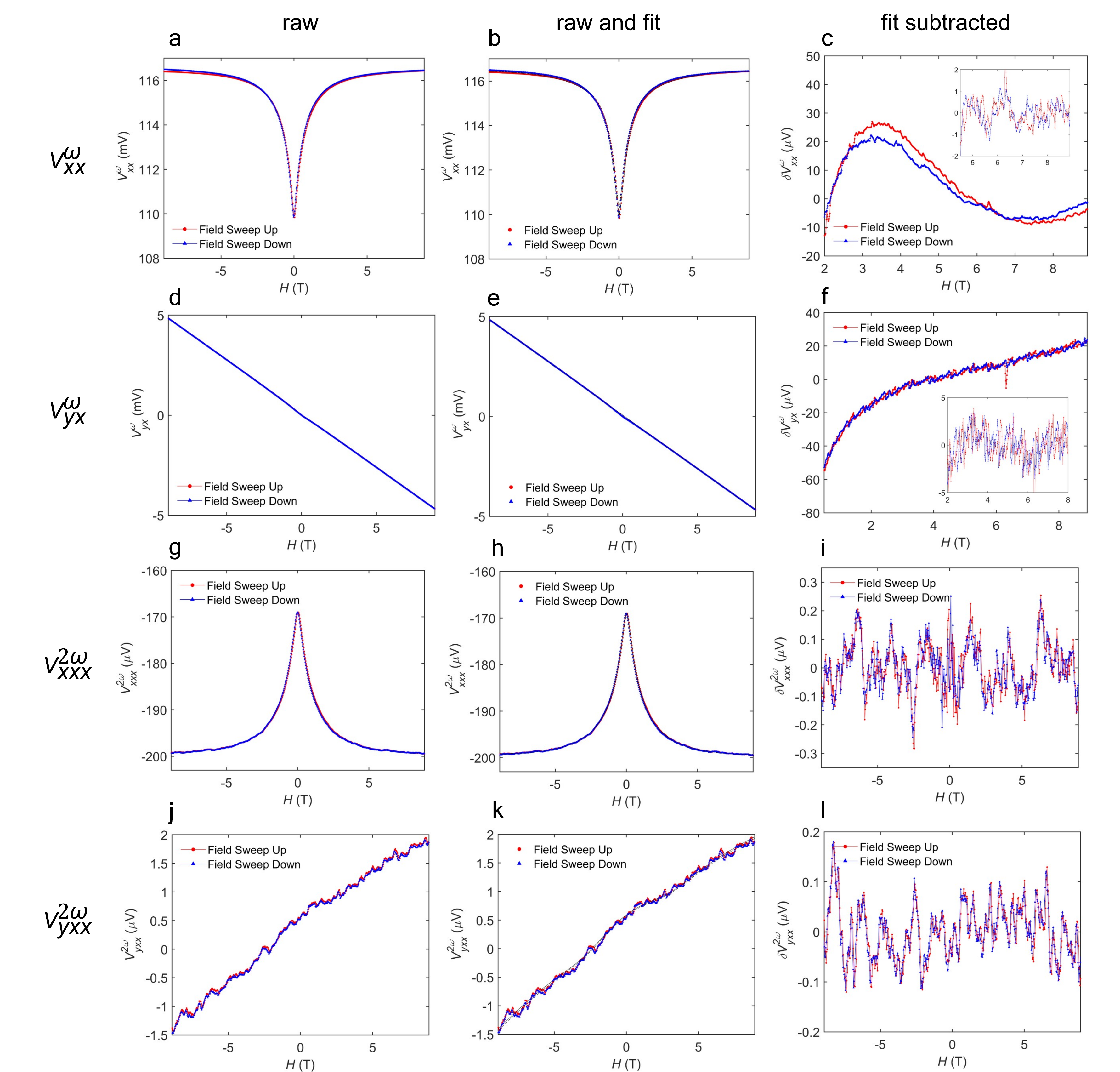}% Here is how to import EPS art
\phantomcaption\label{fig:KTO111fluctchecks_a}
\phantomcaption\label{fig:KTO111fluctchecks_b}
\phantomcaption\label{fig:KTO111fluctchecks_c}
\phantomcaption\label{fig:KTO111fluctchecks_d}
\phantomcaption\label{fig:KTO111fluctchecks_e}
\phantomcaption\label{fig:KTO111fluctchecks_f}
\phantomcaption\label{fig:KTO111fluctchecks_g}
\phantomcaption\label{fig:KTO111fluctchecks_h}
\phantomcaption\label{fig:KTO111fluctchecks_i}
\phantomcaption\label{fig:KTO111fluctchecks_j}
\phantomcaption\label{fig:KTO111fluctchecks_k}
\phantomcaption\label{fig:KTO111fluctchecks_l}
\end{subcaptiongroup}
\captionsetup{subrefformat=parens}
\caption[Nonlinear fluctuations in out-of-plane magnetic field in $\mathrm{LaAlO_3/KTaO_3(111)}$]{\label{fig:KTO111fluctchecks} \justifying \textbf{Nonlinear fluctuations in out-of-plane magnetic field in $\mathbf{LaAlO_3/KTaO_3(111)}$}. The fitting procedure is demonstrated for each of the measured responses according to Eq.~\ref{eq:fitFluctEq}. The results are shown for $V_{xx}^{\omega}$ in \subref*{fig:KTO111fluctchecks_a}--\subref*{fig:KTO111fluctchecks_c}, for $V_{yx}^{\omega}$ in \subref*{fig:KTO111fluctchecks_d}--\subref*{fig:KTO111fluctchecks_f}, for $V_{xxx}^{2\omega}$ in \subref*{fig:KTO111fluctchecks_g}--\subref*{fig:KTO111fluctchecks_i}, and for $V_{yxx}^{2\omega}$ in \subref*{fig:KTO111fluctchecks_j}--\subref*{fig:KTO111fluctchecks_l}. In panels \subref*{fig:KTO111fluctchecks_c} and \subref*{fig:KTO111fluctchecks_f}, we demonstrate that the fluctuations are not detected in the first-harmonic signals. To further illustrate this point, after subtracting the primary fitting function, we subtract an additional second-degree polynomial from the data in a narrow field range (see insets), where only non-reproducible noise is observed. In contrast, reproducible fluctuations are conspicuous in the second-harmonic transverse response (panel \subref*{fig:KTO111fluctchecks_l}). They are also observable in the second-harmonic longitudinal response (panel \subref*{fig:KTO111fluctchecks_i}). We attribute the longitudinal second-harmonic fluctuations to the admixture of the transverse signal into the longitudinal measurement. For the transverse responses, the raw data are presented after correcting for mixing from the $x$-direction. When fitting the $V_{xxx}^{2\omega}$ data, we allow for a small field shift of $\sim 20$~mT to account for the slight field hysteresis of the superconducting magnet. Measurements were taken at $I^\omega=50~\mathrm{\mu A}$, $V_{\mathrm{G}}=24$~V, and $T=2.75$~K.
}
\end{figure*}

To accurately isolate the fluctuations in the second-harmonic Hall measurements and to demonstrate their absence in the first harmonic, a rigorous background subtraction procedure was employed. First, we accounted for the longitudinal voltage ($V_{xx}$) mixing into the transverse Hall signal ($V_{yx}$), which arises from geometric contact misalignment. Because the first-harmonic Hall voltage is theoretically expected to vanish at zero magnetic field for a non-magnetic sample, a geometric mixing factor was determined as $\alpha = V_{yx}^{\omega} / V_{xx}^{\omega}$ near $H = 0$. The corrected, pure Hall responses for both harmonics were then calculated by subtracting this longitudinal contribution:
\begin{equation}
    V_{yx, \mathrm{corr}}^{n\omega} = V_{yx}^{n\omega} - \alpha V_{xx}^{n\omega}, \qquad n \in \{1, 2\}.
    \label{eq:offsetEq}
\end{equation}
This correction procedure is demonstrated in Figs.~\ref{fig:KTO111fluctchecks} and \ref{fig:KTO111NLHE}--\ref{fig:KTO111VIVG}. 

For the first-harmonic response, the longitudinal signal was fitted using the weak antilocalization theory described in Sec.~\ref{sec:WAL}, while the Hall signal was evaluated using a standard linear fit. In the absence of a simple theoretical model for the nonlinear terms, the macroscopic background of the second-harmonic signal, $V_{yxx, \mathrm{corr}}^{2\omega}(H)$, was instead modeled using the following empirical function:
\begin{equation}
\label{eq:fitEq}
    V_{\mathrm{fit}}(H) = \frac{d H^2}{\left[1 + (H/e)^f\right]^g} + c H^2 + b H + a.
\end{equation}
To ensure robust convergence of the nonlinear fitting algorithm, initial parameter guesses were systematically derived from the asymptotic behavior of the experimental data. The parameters $a$, $b$, and $c$ were estimated from the $y$-intercept and a general parabolic fit. The coefficient $d$ was isolated by examining the low-field limit ($H \to 0$), where the dominant quadratic term is approximately $(d+c)H^2$. The parameter $e$ was estimated based on the high-field limit ($H/e \gg 1$), and the exponents $f$ and $g$ were initialized to the trivial values of 2 and 1, respectively.

Finally, the mesoscopic fluctuations of interest, $\delta V_{yxx}^{2\omega}(H)$, were extracted by subtracting the optimized macroscopic background model from the corrected experimental data:
\begin{equation}
\label{eq:fitFluctEq}
    \delta V_{yxx}^{2\omega}(H) = V_{yxx, \mathrm{corr}}^{2\omega}(H) - V_{\mathrm{fit}}(H).
\end{equation}
A similar fitting function was used to extract the fluctuations in the longitudinal second-harmonic response, $\delta V_{xxx}^{2\omega}(H)$. This final background subtraction is demonstrated in Fig.~\ref{fig:KTO111fluctchecks} for all four measured responses.  

For the gate-voltage-dependent measurements, we applied the same contact misalignment correction according to Eq.~(\ref{eq:offsetEq}). We then utilized a simple third-degree polynomial background fit to isolate the voltage fluctuations, $\delta V_{yxx}^{2\omega}(V_{\mathrm{G}})$, and the corresponding gate-tuned conductivity fluctuations, $\delta \sigma_{yxx}(V_{\mathrm{G}})$, as shown in Figs.~1j--k of the main text.

\subsection{Magnetic correlation length}
To analyze the in-field nonlinear fluctuation patterns, we calculate the autocorrelation function:
\begin{equation}
    F(\Delta H) = \langle \delta\sigma_{yxx}(H)\delta\sigma_{yxx}(H+\Delta H) \rangle.
\end{equation}
The root-mean-square (rms) amplitude of the fluctuations is given by $\delta\sigma_{yxx}^{\mathrm{rms}}=\sqrt{F(0)}$. We define the magnetic correlation field, $H_c$, as the half-width at half-maximum, such that $F(H_c)=\frac{1}{2}F(0)$. To extract a characteristic magnetic correlation length $l_f$ associated with the fluctuations, we use the relation $l_f=\sqrt{\phi_0/H_c}$, where $\phi_0=h/e$ is the single-electron magnetic flux quantum. 

\begin{figure*}
\begin{subcaptiongroup}
\includegraphics[width=.7\columnwidth]{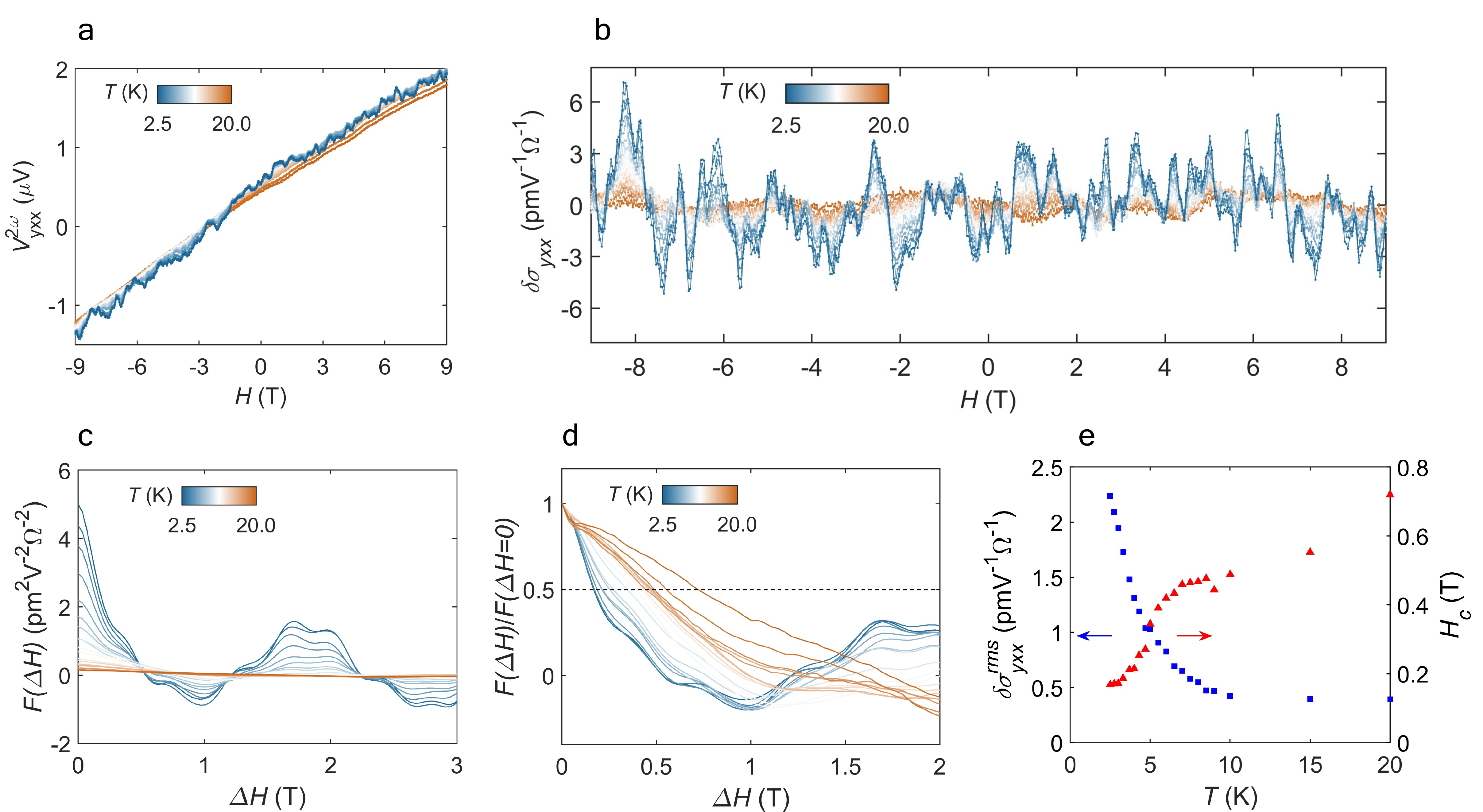}% Here is how to import EPS art
\phantomcaption\label{fig:KTO111ACFH_a}
\phantomcaption\label{fig:KTO111ACFH_b}
\phantomcaption\label{fig:KTO111ACFH_c}
\phantomcaption\label{fig:KTO111ACFH_d}
\phantomcaption\label{fig:KTO111ACFH_e}
\end{subcaptiongroup}
\captionsetup{subrefformat=parens}
\caption[Temperature dependence of nonlinear Hall fluctuations in $\mathrm{LaAlO_3/KTaO_3}$(111)]{\label{fig:KTO111ACFH} \justifying \textbf{Temperature dependence of nonlinear Hall fluctuations in $\mathbf{LaAlO_3/KTaO_3}$(111)}. 
\subref*{fig:KTO111ACFH_a} Second-harmonic Hall voltage $V_{yxx}^{2\omega}$ as a function of the magnetic field $H$ at various temperatures. 
\subref*{fig:KTO111ACFH_b} Nonlinear Hall conductivity fluctuations $\delta\sigma_{yxx}$ as a function of $H$. 
\subref*{fig:KTO111ACFH_c} Autocorrelation function $F(\Delta H)$ of the conductivity fluctuations as a function of the magnetic field shift $\Delta H$. 
\subref*{fig:KTO111ACFH_d} Normalized autocorrelation function $F(\Delta H)/F(0)$ as a function of $\Delta H$. 
\subref*{fig:KTO111ACFH_e} Temperature dependence of the extracted root-mean-square amplitude $\delta\sigma_{yxx}^{\mathrm{rms}}$ and the correlation field $H_c$. The extracted quantities are used in Figs.~2e--g of the main text (measured during a separate cooldown from the data shown in Figs.~1h and 1i). All measurements were taken at a gate voltage of $V_{\mathrm{G}} = 24$~V.
}
\end{figure*}

As shown in Fig.~\ref{fig:KTO111ACFH}, we calculate $F(\Delta H)$ from the experimental data at various temperatures. From these profiles, we extract the temperature dependence of $\delta\sigma_{yxx}^{\mathrm{rms}}$, $H_c$, and, consequently, $l_f$.

\section{Additional data and discussions}
\subsection{Weak antilocalization measurements}\label{sec:WAL}
The linear magnetoconductance data are analyzed using the Maekawa-Fukuyama (MF) model \cite{maekawa1981magnetoresistance, caviglia2010tunable}:
\begin{equation}
\label{eq:WAL}
\begin{split}
\frac{\Delta\sigma_{xx}(H)}{\sigma_\mathrm{Q}} = &-\frac{1}{2}\psi\left(\frac{1}{2}+\frac{H_{i}}{H}\right) + \frac{1}{2}\ln\left(\frac{H_{i}}{H}\right) \\
&+ \psi\left(\frac{1}{2}+\frac{H_{i}+H_{so}}{H}\right) - \ln\left(\frac{H_{i}+H_{so}}{H}\right) \\
&+ \frac{1}{2}\psi\left(\frac{1}{2}+\frac{H_{i}+2H_{so}}{H}\right) - \frac{1}{2}\ln\left(\frac{H_{i}+2H_{so}}{H}\right) \\
&- A_{k}\frac{\sigma_{xx}(0)}{\sigma_\mathrm{Q}}\frac{H^{2}}{1+CH^{2}},
\end{split}
\end{equation}
where $\sigma_\mathrm{Q}=e^2/(\pi h)$ and $\psi$ is the digamma function. The last term describes the classical orbital magnetoconductance with empirical fitting parameters $A_{k}$ and $C$. The parameters $H_{i,so} = \hbar/(4eD\tau_{i,so})$ are the characteristic phase-coherence (inelastic) and spin-orbit effective fields, respectively, and $D = \pi\hbar^{2}\sigma_{xx}(0)/(e^2m^{*})$ is the diffusion constant. Here, we assume an electron effective mass of $m^{*} = 0.23m_{e}$ \cite{vicente2021spin}. 

The characteristic length scales for inelastic and spin-orbit scattering are extracted according to $l_{\phi} = \sqrt{D\tau_{i}}$ and $l_{so} = \sqrt{D\tau_{so}}$. The effective Rashba spin-orbit coupling parameter can be evaluated as $\alpha_{\mathrm{R}} = \hbar^{2}/(2m^{*}l_{so})$. We compare these extracted time scales with the transport time $\tau$ and the associated mean free path $l_\mathrm{tr}$, evaluated from the Drude relation $\sigma_{xx} = n_\mathrm{H}e^{2}\tau/m^{*}$, where the electron density $n_\mathrm{H}$ is extracted from Hall effect measurements. The effective spin-orbit energy is then calculated according to $\Delta_{so} = 2\alpha_\mathrm{R}k_\mathrm{F}$, where the Fermi wavevector is $k_\mathrm{F} = \sqrt{2\pi n_\mathrm{H}}$. 

\begin{figure*}
\begin{subcaptiongroup}
\includegraphics[width=1\columnwidth]{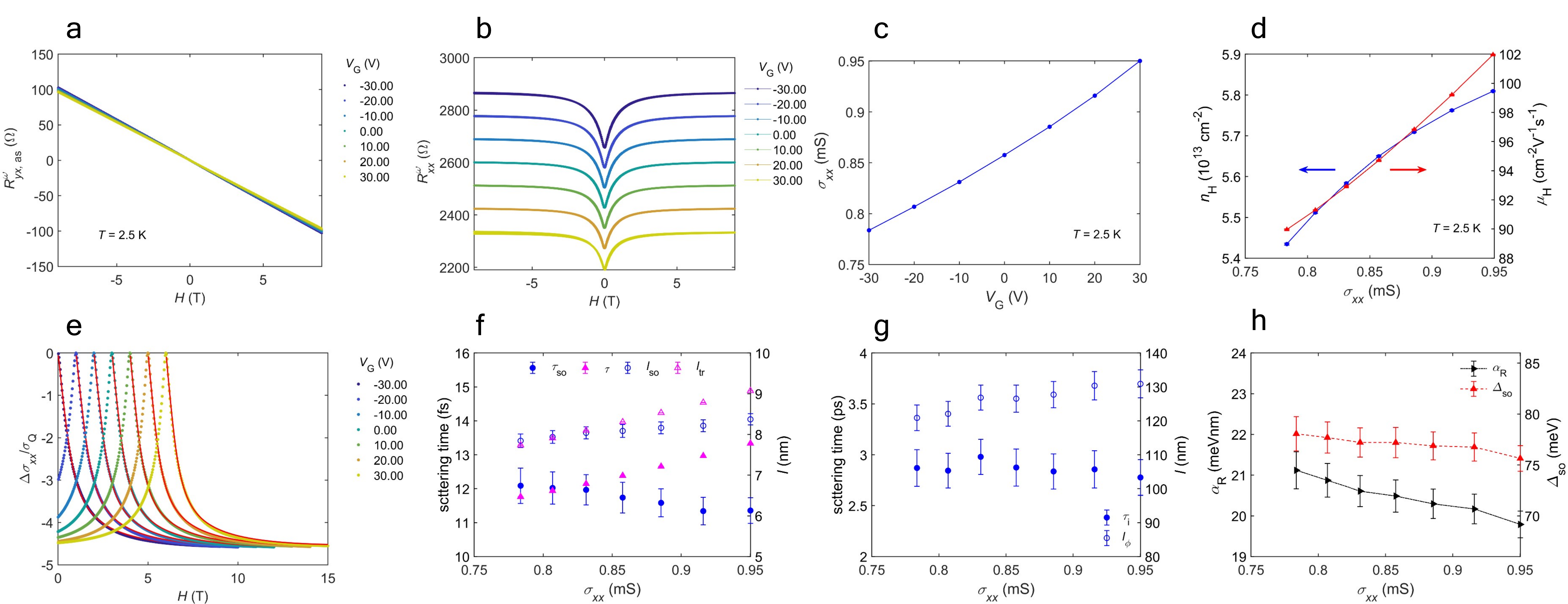}% Here is how to import EPS art
\phantomcaption\label{fig:111WALVg_a}
\phantomcaption\label{fig:111WALVg_b}
\phantomcaption\label{fig:111WALVg_c}
\phantomcaption\label{fig:111WALVg_d}
\phantomcaption\label{fig:111WALVg_e}
\phantomcaption\label{fig:111WALVg_f}
\phantomcaption\label{fig:111WALVg_g}
\phantomcaption\label{fig:111WALVg_h}
\end{subcaptiongroup}
\captionsetup{subrefformat=parens}
\caption[Gate-tuned weak antilocalization in $\mathrm{LaAlO_3/KTaO_3(111)}$]{\label{fig:111WALVg} \justifying \textbf{Gate-tuned weak antilocalization in $\mathbf{LaAlO_3/KTaO_3(111)}$}. \subref*{fig:111WALVg_a} The anti-symmetrized linear Hall resistance $R_{yx, as}^{\omega}$ as a function of $H$ at selected $V_\mathrm{G}$. \subref*{fig:111WALVg_b} Raw data of $R_{xx}^{\omega}$ as a function of $H$ at selected $V_\mathrm{G}$. \subref*{fig:111WALVg_c} $\sigma_{xx}$ obtained at $H=0$ as a function of $V_\mathrm{G}$. \subref*{fig:111WALVg_d} Extracted Hall electron density $n_\mathrm{H}$ and mobility $\mu_\mathrm{H}$ as a function of $\sigma_{xx}$. \subref*{fig:111WALVg_e} Gate-tuned magnetoconductance curves alongside  Maekawa-Fukuyama fits according to Eq. \ref{eq:WAL} (red solid lines). Consecutive curves are shifted horizontally by $1$ T for clarity.  \subref*{fig:111WALVg_f}-\subref*{fig:111WALVg_g} Extracted time scales $\tau_{so}$, $\tau_{i}$, and $\tau$ and their corresponding length scales $l_{so}$, $l_\phi$, and $l_\mathrm{tr}$ as a function of $\sigma_{xx}$. \subref*{fig:111WALVg_h} Extracted effective Rashba spin-orbit coupling parameter $\alpha_\mathrm{R}$ and spin-orbit coupling energy $\Delta_{so}$ as a function of $\sigma_{xx}$. Measurements were taken at $T=2.5$ K.
}
\end{figure*}

\begin{figure*}
\begin{subcaptiongroup}
\includegraphics[width=.8\columnwidth]{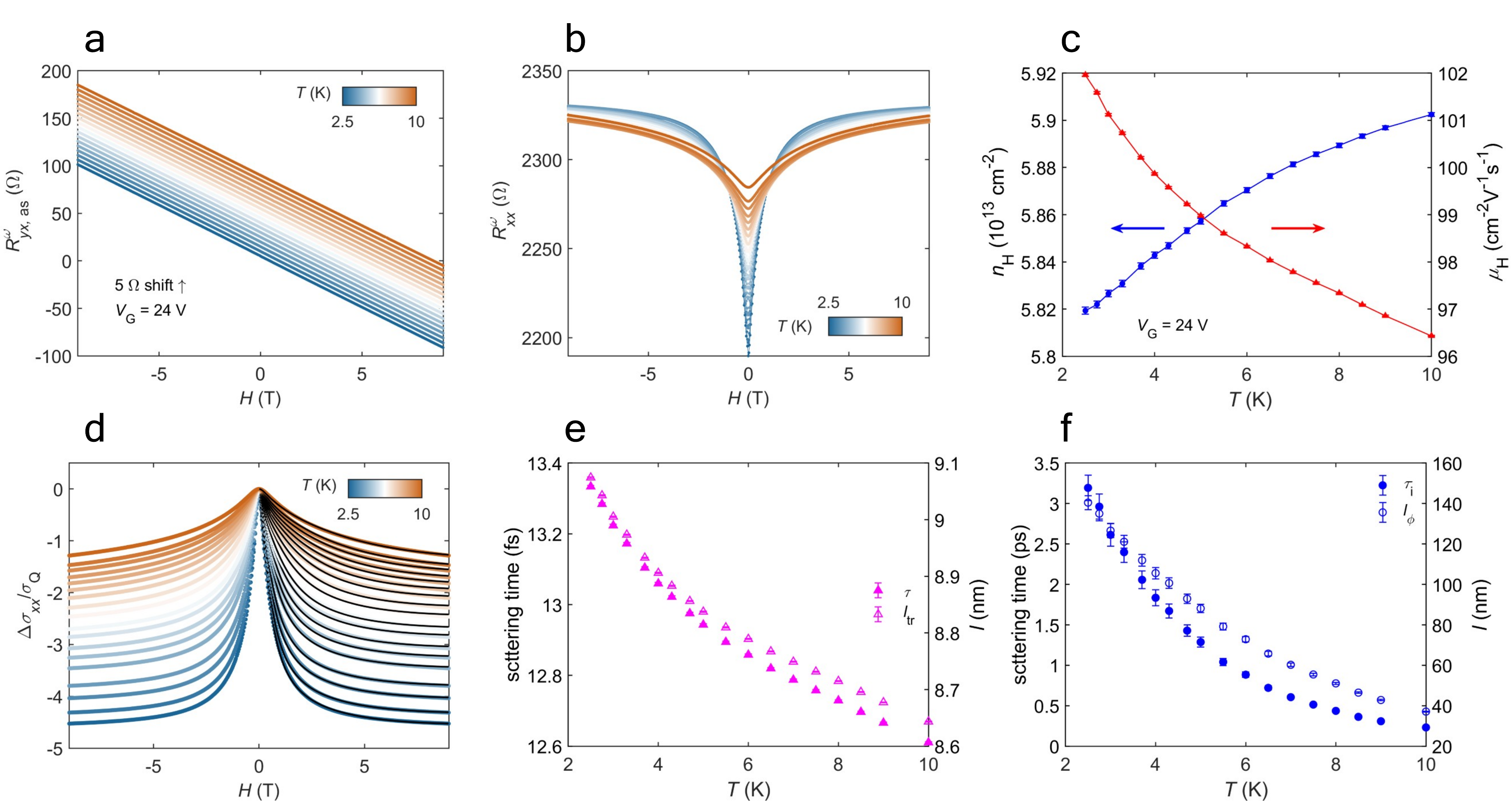}% Here is how to import EPS art
\phantomcaption\label{fig:111WALT_a}
\phantomcaption\label{fig:111WALT_b}
\phantomcaption\label{fig:111WALT_c}
\phantomcaption\label{fig:111WALT_d}
\phantomcaption\label{fig:111WALT_e}
\phantomcaption\label{fig:111WALT_f}

\end{subcaptiongroup}
\captionsetup{subrefformat=parens}
\caption[Temperature dependent weak antilocalization in $\mathrm{LaAlO_3/KTaO_3(111)}$]{\label{fig:111WALT} \justifying \textbf{Temperature dependent weak antilocalization in $\mathbf{LaAlO_3/KTaO_3(111)}$}. \subref*{fig:111WALT_a} The anti-symmetrized linear Hall resistance $R_{yx, as}^{\omega}$ as a function of $H$ at various $T$. Consecutive curves are shifted vertically by $5\ \Omega$ for clarity. \subref*{fig:111WALT_b} Raw data of $R_{xx}^{\omega}$ as a function of $H$ at various $T$. \subref*{fig:111WALT_c} Extracted Hall electron density $n_\mathrm{H}$ and mobility $\mu_\mathrm{H}$ as a function of $T$. \subref*{fig:111WALT_d} Temperature dependent magnetoconductance curves alongside  Maekawa-Fukuyama fits according to Eq. \ref{eq:WAL} (black solid lines). \subref*{fig:111WALT_e}-\subref*{fig:111WALT_f} Extracted time scales $\tau$, $\tau_{i}$ and their corresponding length scales $l_{tr}$, $l_\phi$ as a function of $T$. The effective spin-orbit field has been extracted from the WAL fit at $T=2.5$ K, $H_{so}=2.4\pm0.1$ T, and it is assumed to be temperature independent. Measurements were taken at $V_\mathrm{G}=24$ V.  
}
\end{figure*}

The results for the weak antilocalization measurements under gate-voltage and temperature variations are summarized in Figs.~\ref{fig:111WALVg} and \ref{fig:111WALT}, respectively. As shown in Fig.~\ref{fig:111WALVg_h}, we obtain a typical Rashba spin-orbit coupling energy of $\Delta_{so} \simeq 75$~meV, which is in excellent agreement with previous studies \cite{Zhang2019unusual, gan2023light}. This energy decreases as a function of $\sigma_{xx}$, which in turn is increased by raising $V_\mathrm{G}$. For simplicity in the weak antilocalization analysis shown in Fig.~\ref{fig:111WALT}, we assume a temperature-independent spin-orbit coupling field $H_{so}$ over the narrow temperature range of $2.5\text{--}10$~K.

\subsection{Control experiments}
\begin{figure*}
\begin{subcaptiongroup}
\includegraphics[width=1\columnwidth]{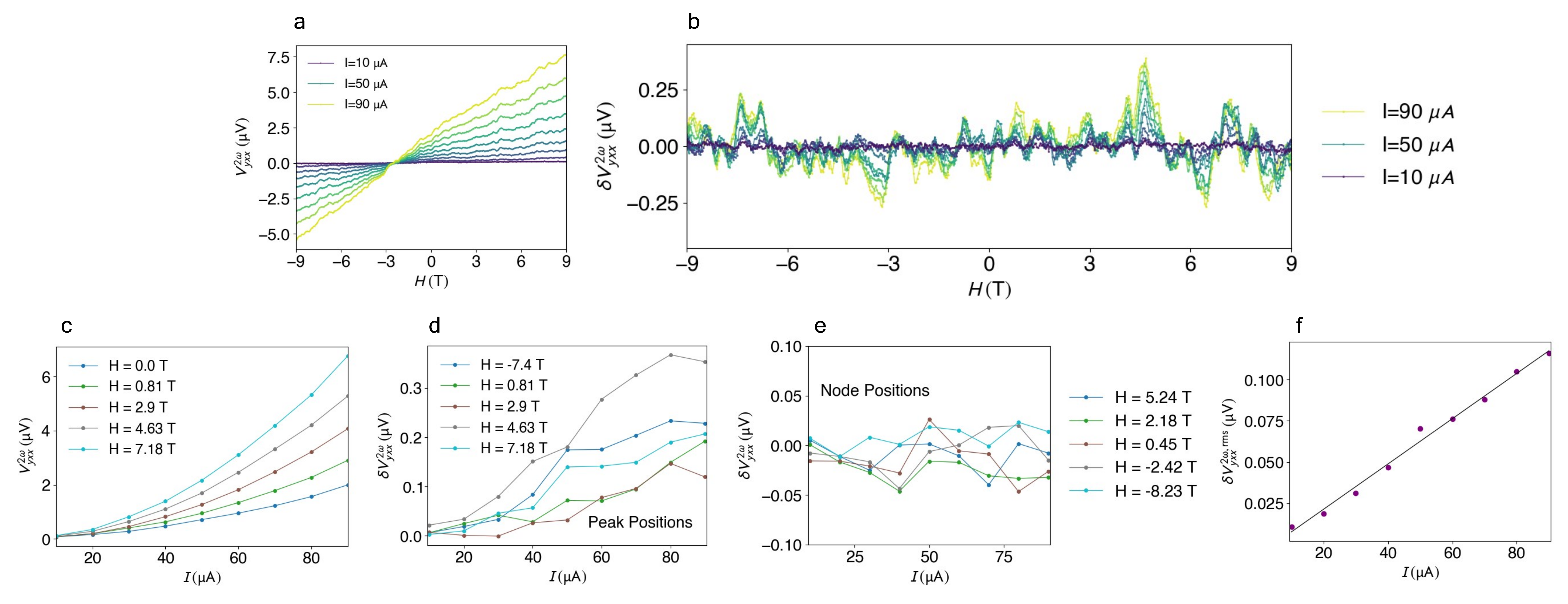}% Here is how to import EPS art
\phantomcaption\label{fig:Curr_dep_a}
\phantomcaption\label{fig:Curr_dep_b}
\phantomcaption\label{fig:Curr_dep_c}
\phantomcaption\label{fig:Curr_dep_d}
\phantomcaption\label{fig:Curr_dep_e}
\phantomcaption\label{fig:Curr_dep_f}
\phantomcaption\label{fig:Curr_dep_g}
\end{subcaptiongroup}
\captionsetup{subrefformat=parens}
\caption[Current dependence of fluctuations in $\mathrm{LaAlO_3/KTaO_3(111)}$]{\label{fig:HE_SC} \justifying \textbf{Current dependence of fluctuations in $\mathbf{LaAlO_3/KTaO_3(111)}$}.
\subref*{fig:Curr_dep_a} $V_{yxx}^{2\omega}$ as a function of the magnetic field $H$, measured at various driving currents $I^{\omega}$, demonstrating the nonlinear Hall effect. 
\subref*{fig:Curr_dep_b} Fluctuations $\delta V_{yxx}^{2\omega}$ as a function of $H$ for various $I^{\omega}$. 
\subref*{fig:Curr_dep_c} $V_{yxx}^{2\omega}$ as a function of $I^{\omega}$ at different magnetic fields, highlighting the parabolic dependence of the nonlinear transverse voltage. 
\subref*{fig:Curr_dep_d} $\delta V_{yxx}^{2\omega}$ at different peak positions plotted as a function of $I^{\omega}$. 
\subref*{fig:Curr_dep_e} $\delta V_{yxx}^{2\omega}$ at node positions as a function of $I^{\omega}$. The field values at which the fluctuations vanish remain constant with increasing current, further ruling out Joule heating as a possible origin. 
\subref*{fig:Curr_dep_f} The rms amplitude of the fluctuations $\delta V_{yxx}^{2\omega,\,rms}$ as a function of $I^{\omega}$, exhibiting the expected linear dependence (solid line).}
\end{figure*}

\begin{figure*}
\begin{subcaptiongroup}
\includegraphics[width=.7\columnwidth]{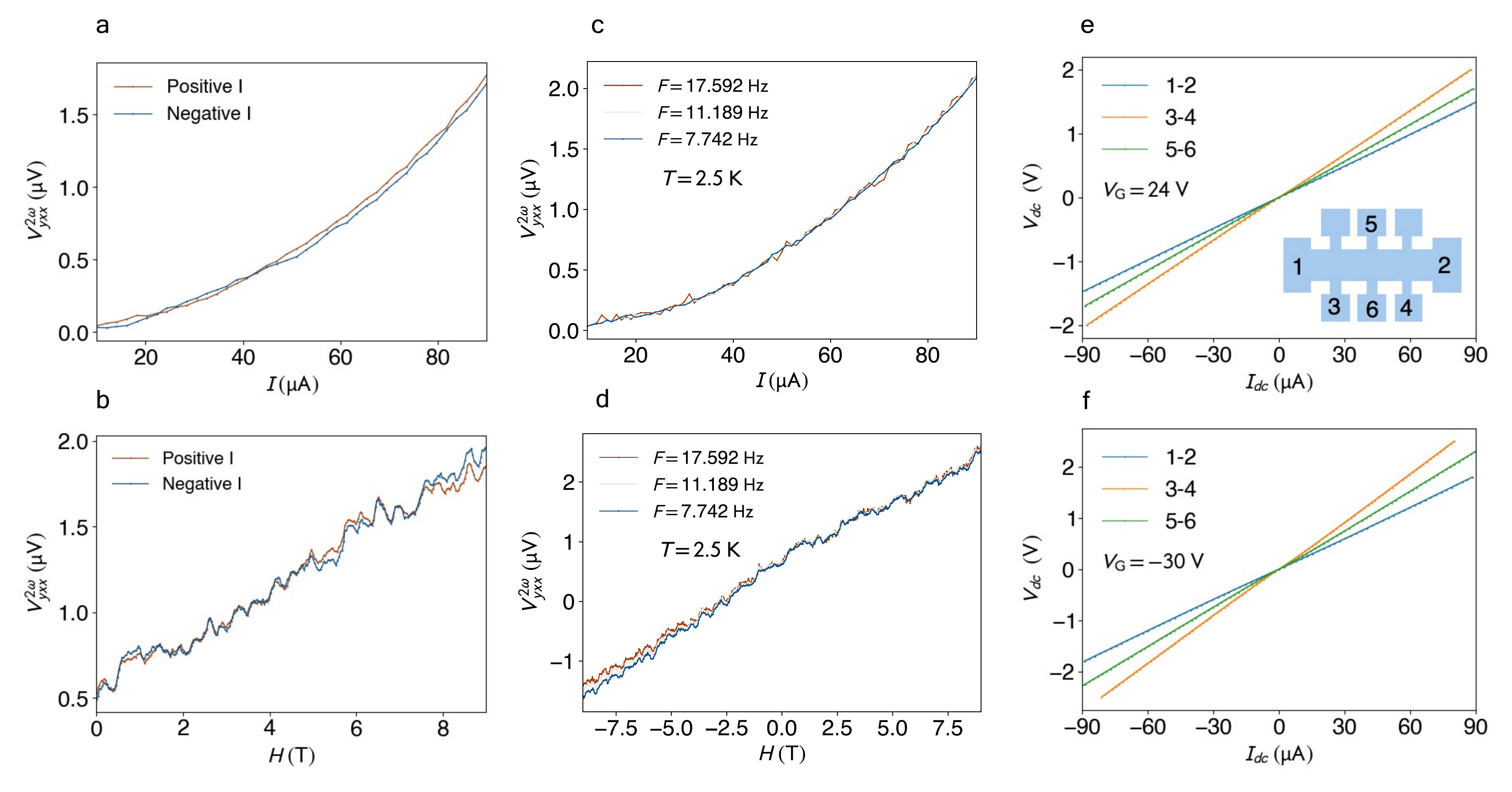}% Here is how to import EPS art
\phantomcaption\label{fig:Curr_direc_a}
\phantomcaption\label{fig:Curr_direc_b}
\phantomcaption\label{fig:Freq_dep_c}
\phantomcaption\label{fig:Freq_dep_d}
\phantomcaption\label{fig:Contact_dep_e}
\phantomcaption\label{fig:Contact_dep_f}
\end{subcaptiongroup}
\captionsetup{subrefformat=parens}
\caption[Nonreciprocal response, frequency independence, and contact characterization in $\mathrm{LaAlO_3/KTaO_3(111)}$]{\label{fig:Freq_dep} \justifying \textbf{Nonreciprocal response, frequency independence, and contact characterization in $\mathbf{LaAlO_3/KTaO_3(111)}$}.
\subref*{fig:Curr_direc_a} $V_{yxx}^{2\omega}$ as a function of the driving current $I^{\omega}$ at zero magnetic field, confirming nonreciprocal response between the positive and negative current measurements. For the negative-current measurements, the voltage probes were interchanged, and the sign of the measured voltage was inverted for direct comparison. 
\subref*{fig:Curr_direc_b} $V_{yxx}^{2\omega}$ as a function of $H$ (same conditions as in \subref*{fig:Curr_direc_a}), demonstrating that positive and negative currents produce identical fluctuation patterns in the raw data.
\subref*{fig:Freq_dep_c} $V_{yxx}^{2\omega}$ as a function of $I^{\omega}$ at zero magnetic field for various excitation frequencies, demonstrating that the signal is independent of the measurement frequency. 
\subref*{fig:Freq_dep_d} $V_{yxx}^{2\omega}$ as a function of $H$, confirming that the raw data fluctuations are also independent of the measurement frequency. 
\subref*{fig:Contact_dep_e} Two-probe measurements of the DC voltage $V_{dc}$ as a function of the applied current $I_{dc}$ for various contact configurations. The linear characteristics confirm that the contacts are Ohmic. Measurements were taken at $V_\mathrm{G}=24$ V and $T=2.5$ K.
\subref*{fig:Contact_dep_f} $V_{dc}$ as a function of $I_{dc}$ (same conditions as in \subref*{fig:Contact_dep_e}) at $V_\mathrm{G}=-30$ V.
}
\end{figure*}

To firmly establish the physical origin of the fluctuations, we rule out several extrinsic mechanisms based on the following observations:
\begin{itemize}
\item The fluctuations are highly reproducible across multiple measurements. If they originated from random thermal or instrumental noise, the fluctuation pattern would differ completely with each consecutive magnetic field sweep.
\item The $\delta V_{yxx}^{2\omega}$ versus $H$ traces exhibit distinct nodes where the fluctuation amplitude vanishes (see Fig.~\ref{fig:Curr_dep_b}). If these fluctuations were driven by thermal effects (such as current-induced Joule heating), the magnetic field positions of these nodes would shift with the amplitude of the applied current. However, as demonstrated in Fig.~\ref{fig:Curr_dep_e}, the node positions are entirely independent of the applied current. 
\item Measurements of the dc voltage across various contacts yield perfectly linear $V_{\mathrm{dc}}(I_{\mathrm{dc}})$ curves, as shown in Figs.~\ref{fig:Contact_dep_e}--\ref{fig:Contact_dep_f}. These strictly linear characteristics confirm that the contacts are Ohmic and do not generate a spurious macroscopic nonlinear response.
\end{itemize}

\begin{figure*}
\begin{subcaptiongroup}
\includegraphics[width=.55\columnwidth]{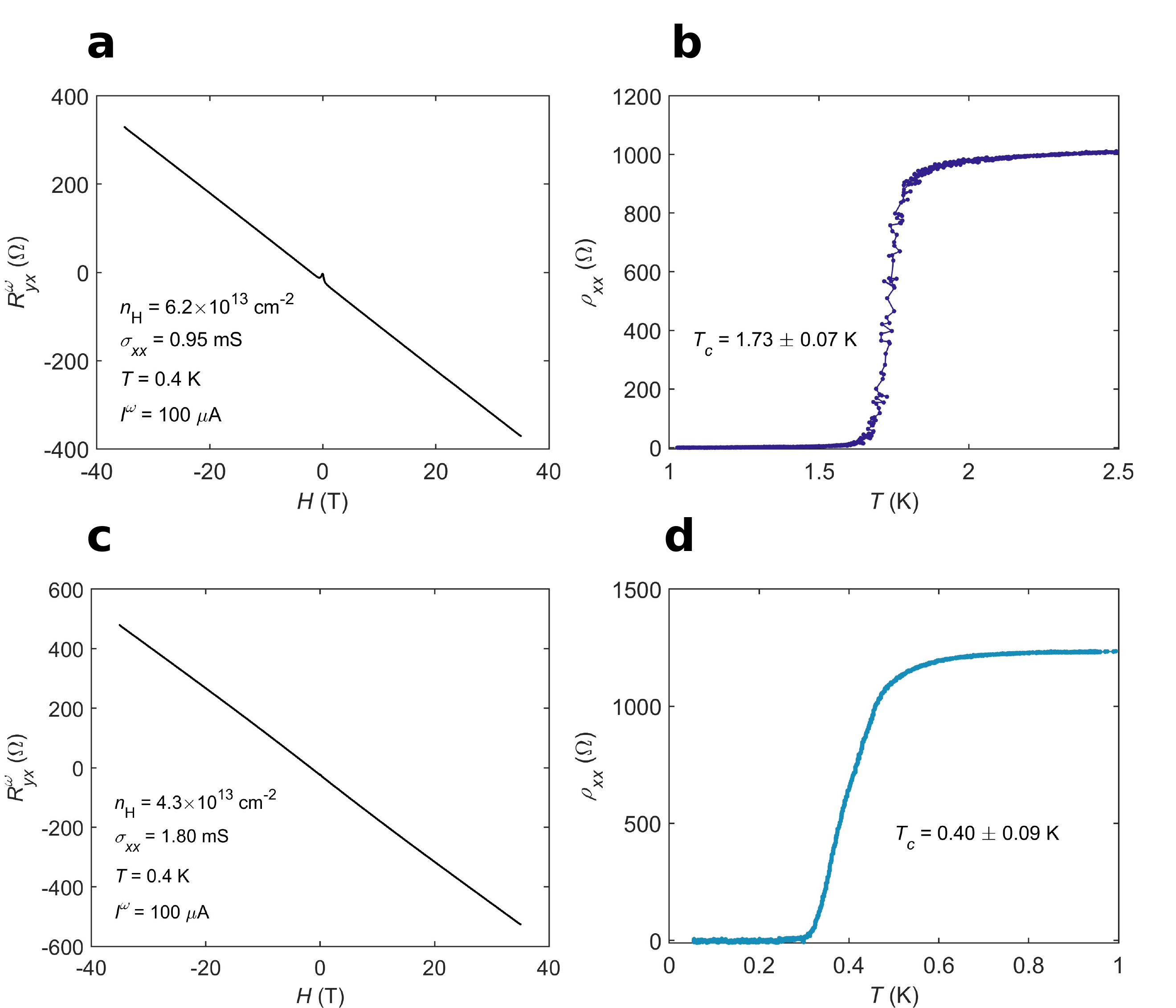}% Here is how to import EPS art
\phantomcaption\label{fig:HE_SC_a}
\phantomcaption\label{fig:HE_SC_b}
\phantomcaption\label{fig:HE_SC_c}
\phantomcaption\label{fig:HE_SC_d}
\end{subcaptiongroup}
\captionsetup{subrefformat=parens}
\caption[Hall effect and superconducting transition in $\mathrm{LaAlO_3/KTaO_3(111)}$ and $\mathrm{LaAlO_3/KTaO_3(110)}$]{\label{fig:HE_SC} \justifying \textbf{Hall effect and superconducting transition in $\mathbf{LaAlO_3/KTaO_3(111)}$ and $\mathbf{LaAlO_3/KTaO_3(110)}$}. \subref*{fig:HE_SC_a} $R^{\omega}_{yx}$ as a function of $H$, showing linear Hall effect up to $H=35$ T in $\mathrm{LaAlO_3/KTaO_3(111)}$. \subref*{fig:HE_SC_b} $\rho_{xx}$ versus $T$, showing a sharp superconducting transition at $T_c=1.73$ K. The high critical temperature is consistent with previous studies \cite{LiuScience2021}, while the sharp transition indicates the sample is highly homogeneous. \subref*{fig:HE_SC_c}-\subref*{fig:HE_SC_d} Linear Hall effect and superconducting transition at $T_c=0.4$ K in $\mathrm{LaAlO_3/KTaO_3(110)}$. The superconducting transition in $\mathrm{LaAlO_3/KTaO_3(110)}$ has been measured for a sample from the same batch and deposited in similar conditions to the sample presented in the main text.}
\end{figure*}

\begin{figure*}
\begin{subcaptiongroup}
\includegraphics[width=.8\columnwidth]{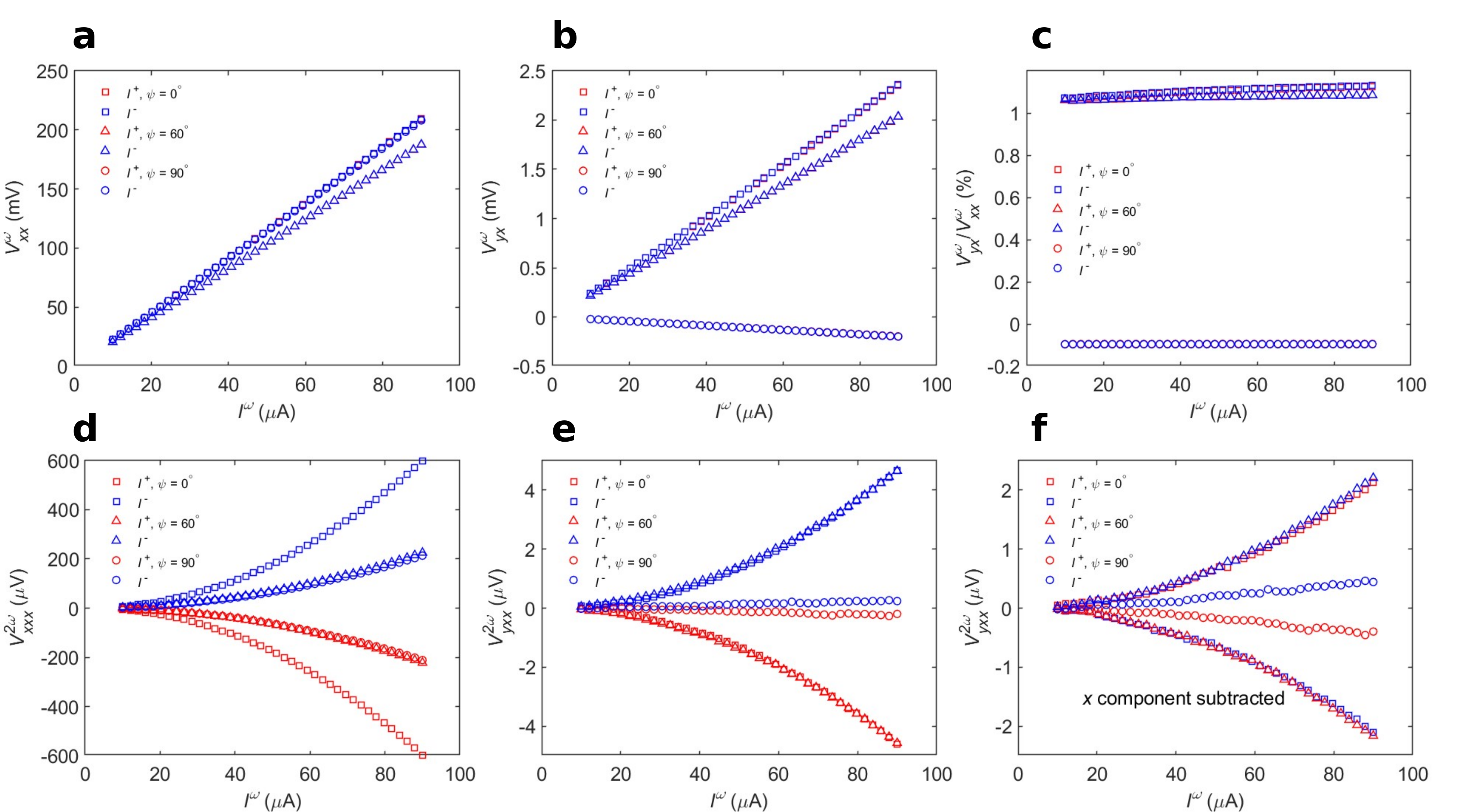}% Here is how to import EPS art
\phantomcaption\label{fig:KTO111NLHE_a}
\phantomcaption\label{fig:KTO111NLHE_b}
\phantomcaption\label{fig:KTO111NLHE_c}
\phantomcaption\label{fig:KTO111NLHE_d}
\phantomcaption\label{fig:KTO111NLHE_e}
\phantomcaption\label{fig:KTO111NLHE_f}
\end{subcaptiongroup}
\captionsetup{subrefformat=parens}
\caption[Nonlinear Hall effect in $\mathrm{LaAlO_3/KTaO_3(111)}$]{\label{fig:KTO111NLHE} \justifying \textbf{Nonlinear Hall effect in $\mathbf{LaAlO_3/KTaO_3(111)}$}. \subref*{fig:KTO111NLHE_a}-\subref*{fig:KTO111NLHE_b} The first harmonic longitudinal and transverse voltages, $V_{xx}^{\omega}$ and $V_{yx}^{\omega}$, as a function of ac current $I^{\omega}$ for three different current injection angles (see Fig. 2f for demonstration). The finite $V_{yx}^{\omega}$ can be attributed to contact misalignment. \subref*{fig:KTO111NLHE_c} The ratio of misalignment $V_{yx}^{\omega}/V_{xx}^{\omega}$ vs $I^{\omega}$. \subref*{fig:KTO111NLHE_d}-\subref*{fig:KTO111NLHE_e} The second harmonic longitudinal and transverse voltages, $V_{xxx}^{2\omega}$ and $V_{yxx}^{2\omega}$, as a function of $I^{\omega}$. \subref*{fig:KTO111NLHE_f} $V_{yxx}^{2\omega}$ vs $I^{\omega}$ after subtracting the contribution of contact misalignment from the $x$ direction. Data are shown for opposite current injection directions while simultaneously exchanging the voltage contact probes. Measurements were taken at $H=0$, $T=2.5$ K, and $V_\mathrm{G}=24$ V.
}
\end{figure*}

\begin{figure*}
\begin{subcaptiongroup}
\includegraphics[width=1\columnwidth]{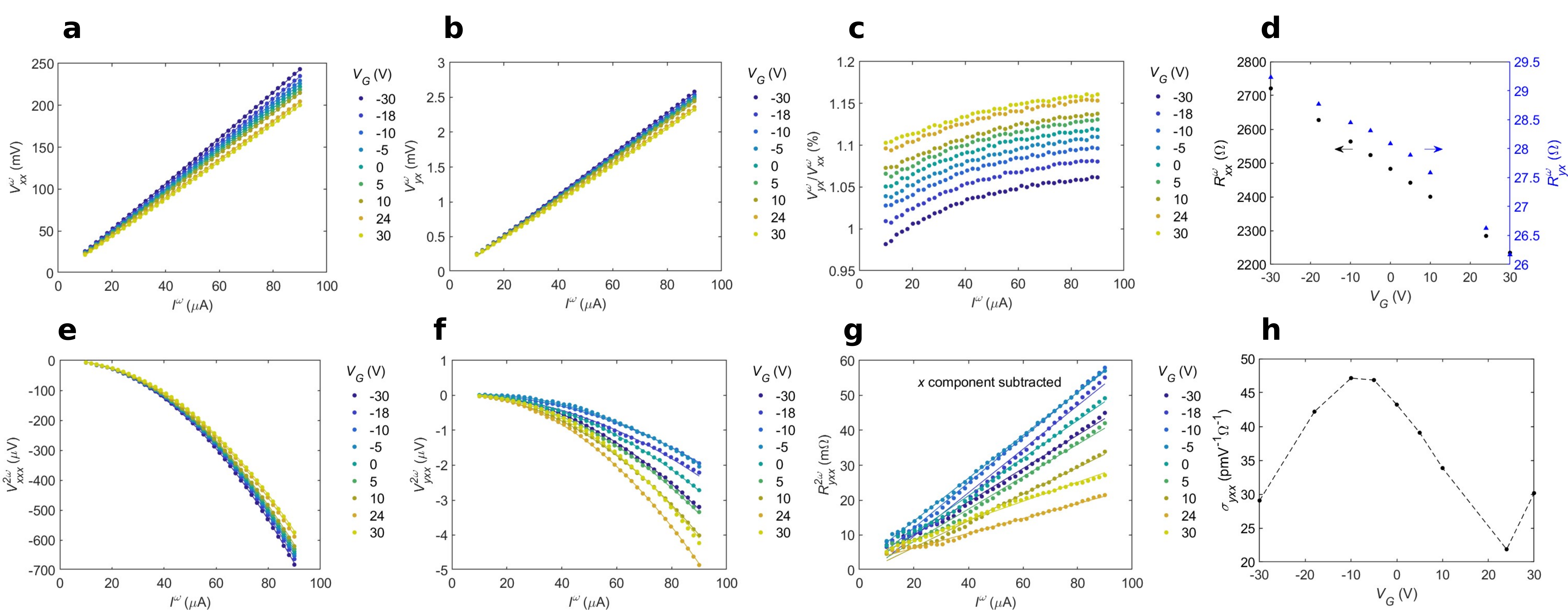}% Here is how to import EPS art
\phantomcaption\label{fig:KTO111VIVG_a}
\phantomcaption\label{fig:KTO111VIVG_b}
\phantomcaption\label{fig:KTO111VIVG_c}
\phantomcaption\label{fig:KTO111VIVG_d}
\phantomcaption\label{fig:KTO111VIVG_e}
\phantomcaption\label{fig:KTO111VIVG_f}
\phantomcaption\label{fig:KTO111VIVG_g}
\phantomcaption\label{fig:KTO111VIVG_h}
\end{subcaptiongroup}
\captionsetup{subrefformat=parens}
\caption[Gate voltage dependence of nonlinear transport in $\mathrm{LaAlO_3/KTaO_3(111)}$]{\label{fig:KTO111VIVG} \justifying \textbf{Gate voltage dependence of nonlinear transport in $\mathbf{LaAlO_3/KTaO_3(111)}$}. \subref*{fig:KTO111VIVG_a}-\subref*{fig:KTO111VIVG_b} The first harmonic longitudinal and transverse voltages, $V_{xx}^{\omega}$ and $V_{yx}^{\omega}$, as a function of ac current $I^{\omega}$ measured at various gate voltages $V_\mathrm{G}$, alongside linear fits (solid lines). The finite $V_{yx}^{\omega}$ can be attributed to contact misalignment. \subref*{fig:KTO111VIVG_c} The ratio of misalignment $V_{yx}^{\omega}/V_{xx}^{\omega}$ vs $I^{\omega}$. \subref*{fig:KTO111VIVG_d} $R_{xx}$ and the ratio $R_{yx}^{\omega}/R_{xx}^{\omega}$ as a function of $V_\mathrm{G}$. \subref*{fig:KTO111VIVG_e}-\subref*{fig:KTO111VIVG_f} The second harmonic longitudinal and transverse voltages, $V_{xxx}^{2\omega}$ and $V_{yxx}^{2\omega}$, as a function of $I^{\omega}$. The solid line represents quadratic fits. \subref*{fig:KTO111VIVG_g} $R_{yxx}^{2\omega}\equiv V_{yxx}^{2\omega}/I^{\omega}$ as a function of $I^{\omega}$ after subtracting the contribution of contact misalignment from the $x$ direction. The solid lines represent linear fits. \subref*{fig:KTO111VIVG_h} The nonlinear Hall conductivity $\sigma_{yxx}$ obtained from the slope of the linear fits in \subref*{fig:KTO111VIVG_g} as a function of $V_\mathrm{G}$. Measurements were taken at $H=0$ and $T=2.5$ K.
}
\end{figure*}

\begin{figure*}
\begin{subcaptiongroup}
\includegraphics[width=.8\columnwidth]{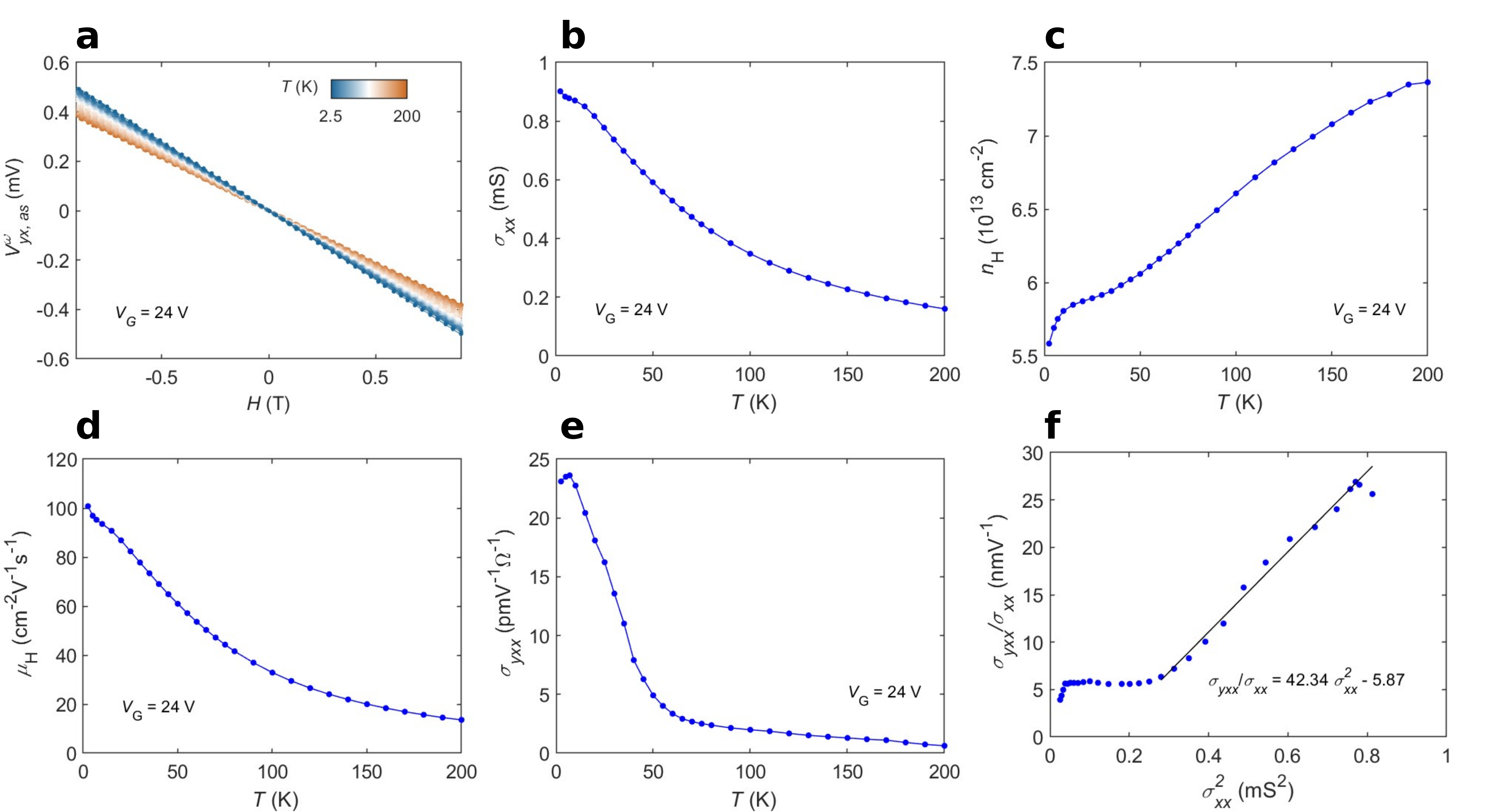}% Here is how to import EPS art
\phantomcaption\label{fig:KTO111Tdepend_a}
\phantomcaption\label{fig:KTO111Tdepend_b}
\phantomcaption\label{fig:KTO111Tdepend_c}
\phantomcaption\label{fig:KTO111Tdepend_d}
\phantomcaption\label{fig:KTO111Tdepend_e}
\phantomcaption\label{fig:KTO111Tdepend_f}

\end{subcaptiongroup}
\captionsetup{subrefformat=parens}
\caption[Temperature dependence of transport properties in $\mathrm{LaAlO_3/KTaO_3(111)}$]{\label{fig:KTO111Tdepend} \justifying \textbf{Temperature dependence of transport properties in $\mathbf{LaAlO_3/KTaO_3(111)}$}. \subref*{fig:KTO111Tdepend_a} The anti-symmetrized linear Hall voltage $V_{yx, as}^{\omega}$ as a function of $H$ at various temperatures. The contribution from the $x$ direction has been subtracted from the data. \subref*{fig:KTO111Tdepend_b} The longitudinal conductivity $\sigma_{xx}$ as a function of $T$. \subref*{fig:KTO111Tdepend_c} The extracted Hall electron density $n_\mathrm{H}$ vs $T$. \subref*{fig:KTO111Tdepend_d} The extracted electron mobility $\mu_\mathrm{H}$ vs $T$. \subref*{fig:KTO111Tdepend_e} The nonlinear Hall conductivity $\sigma_{yxx}$ vs $T$. \subref*{fig:KTO111Tdepend_f} $\sigma_{yxx}/\sigma_{xx}$ as a function of $\sigma_{xx}^{2}$. The solid line represents a linear fit. Measurements were taken at $I^\omega=50\ \mathrm{\mu A}$ and $V_\mathrm{G}=24$ V. Measurements in \subref*{fig:KTO111Tdepend_b} and \subref*{fig:KTO111Tdepend_e} were taken at $H=0$.
}
\end{figure*}

\begin{figure*}
\begin{subcaptiongroup}
\includegraphics[width=1\columnwidth]{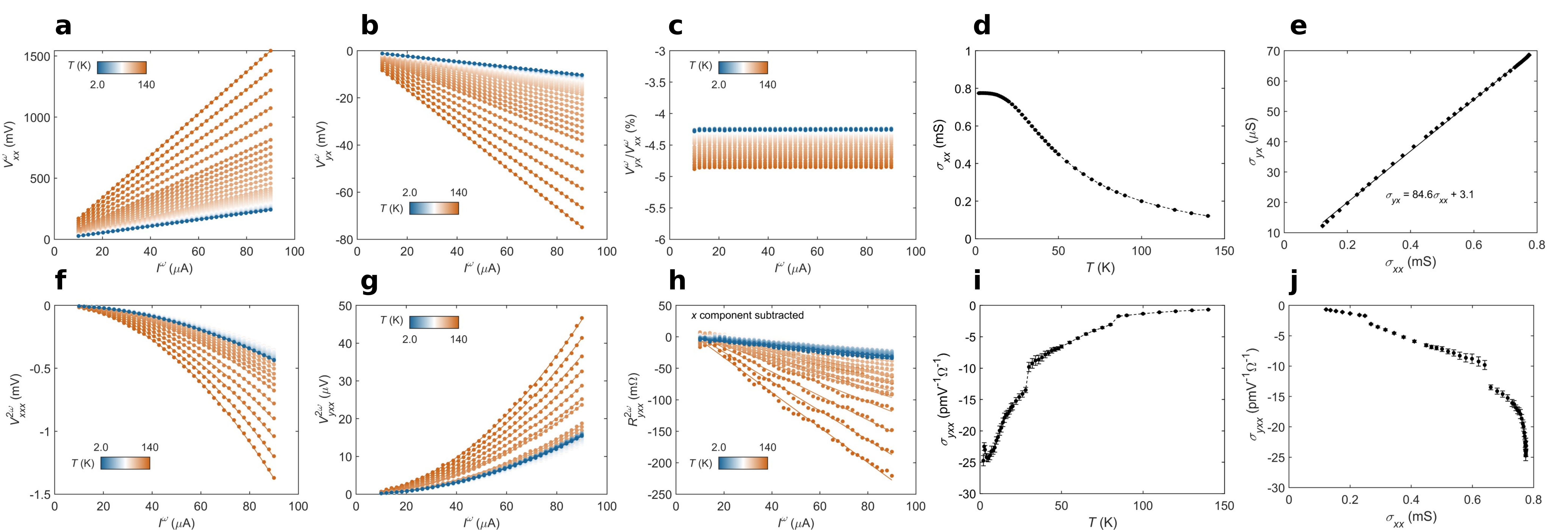}% Here is how to import EPS art
\phantomcaption\label{fig:KTO100transport_a}
\phantomcaption\label{fig:KTO100transport_b}
\phantomcaption\label{fig:KTO100transport_c}
\phantomcaption\label{fig:KTO100transport_d}
\phantomcaption\label{fig:KTO100transport_e}
\phantomcaption\label{fig:KTO100transport_f}
\phantomcaption\label{fig:KTO100transport_g}
\phantomcaption\label{fig:KTO100transport_h}
\phantomcaption\label{fig:KTO100transport_i}
\phantomcaption\label{fig:KTO100transport_j}
\end{subcaptiongroup}
\captionsetup{subrefformat=parens}
\caption[Temperature dependence of nonlinear transport in $\mathrm{LaAlO_3/KTaO_3(100)}$]{\label{fig:KTO100transport} \justifying \textbf{Temperature dependence of nonlinear transport in $\mathbf{LaAlO_3/KTaO_3(100)}$}. \subref*{fig:KTO100transport_a}-\subref*{fig:KTO100transport_b} The first harmonic longitudinal and transverse voltages, $V_{xx}^{\omega}$ and $V_{yx}^{\omega}$, as a function of ac current $I^{\omega}$ measured at various temperatures $T$, alongside linear fits (solid lines). The finite $V_{yx}^{\omega}$ can be attributed to contact misalignment. \subref*{fig:KTO100transport_c} The ratio of misalignment $V_{yx}^{\omega}/V_{xx}^{\omega}$ vs $I^{\omega}$. \subref*{fig:KTO100transport_d} $\sigma_{xx}$ as a function of $T$. \subref*{fig:KTO100transport_e} The zero field $\sigma_{yx}$ vs $\sigma_{xx}$. \subref*{fig:KTO100transport_f}-\subref*{fig:KTO100transport_g} The second harmonic longitudinal and transverse voltages, $V_{xxx}^{2\omega}$ and $V_{yxx}^{2\omega}$, as a function of $I^{\omega}$. The solid lines represent quadratic fits. \subref*{fig:KTO100transport_h} $R_{yxx}^{2\omega}\equiv V_{yxx}^{2\omega}/I^{\omega}$ as a function of $I^{\omega}$ after subtracting the contribution of contact misalignment from the $x$ direction. The solid lines represent linear fits. \subref*{fig:KTO100transport_i} The nonlinear Hall conductivity $\sigma_{yxx}$ obtained from the slope of the linear fits in \subref*{fig:KTO100transport_h} as a function of $T$. \subref*{fig:KTO100transport_j} $\sigma_{yxx}$ as a function of $\sigma_{xx}$. Measurements were taken at $H=0$ and $V_\mathrm{G}=0$.
}
\end{figure*}

\begin{figure*}
\begin{subcaptiongroup}
\includegraphics[width=1\columnwidth]{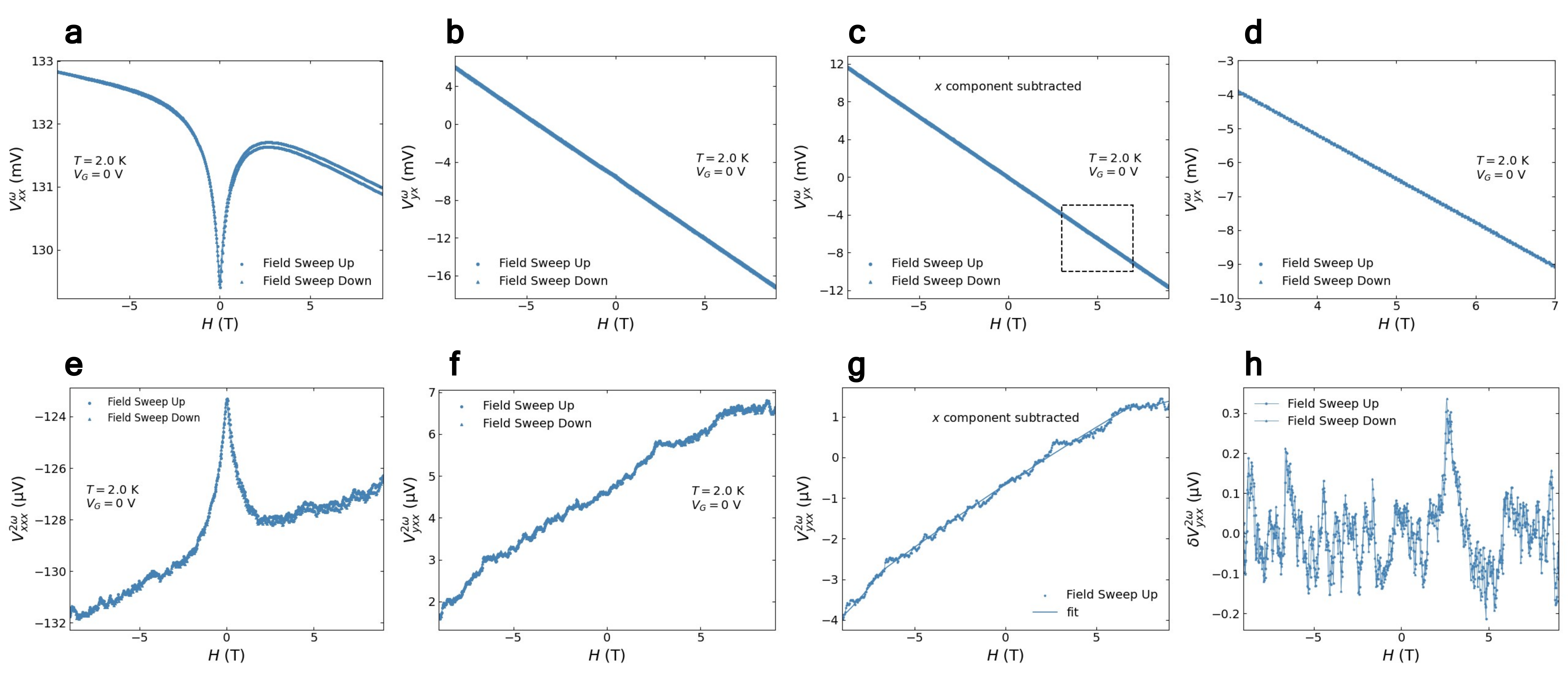}% Here is how to import EPS art
\phantomcaption\label{fig:KTO100fluctH_a}
\phantomcaption\label{fig:KTO100fluctH_b}
\phantomcaption\label{fig:KTO100fluctH_c}
\phantomcaption\label{fig:KTO100fluctH_d}
\phantomcaption\label{fig:KTO100fluctH_e}
\phantomcaption\label{fig:KTO100fluctH_f}
\phantomcaption\label{fig:KTO100fluctH_g}
\phantomcaption\label{fig:KTO100fluctH_h}
\end{subcaptiongroup}
\captionsetup{subrefformat=parens}
\caption[Nonlinear Hall voltage fluctuations in out-of-plane magnetic field in $\mathrm{LaAlO_3/KTaO_3(100)}$]{\label{fig:KTO100fluctH} \justifying \textbf{Nonlinear Hall voltage fluctuations in out-of-plane magnetic field in $\mathbf{LaAlO_3/KTaO_3(100)}$}. \subref*{fig:KTO100fluctH_a}-\subref*{fig:KTO100fluctH_b} First harmonic longitudinal and transverse voltages, $V_{xx}^{\omega}$ and $V_{yx}^{\omega}$, as a function of out-of-plane field $H$. The finite $V_{yx}^{\omega}$ measured at $H=0$ can be due to contact misalignment. \subref*{fig:KTO100fluctH_c} $V_{yx}^{\omega}$ vs $H$ after subtracting the contact misalignment component from the x direction. \subref*{fig:KTO100fluctH_d} Enlarged view of the region indicated by the dashed line box in \subref*{fig:KTO100fluctH_c}. \subref*{fig:KTO100fluctH_e} Second harmonic longitudinal voltage $V^{2\omega}_{xxx}$ vs $H$. The solid line represents a fit to the data. \subref*{fig:KTO100fluctH_f} Second harmonic transverse voltage $V^{2\omega}_{yxx}$ vs $H$. \subref*{fig:KTO100fluctH_g} $V^{2\omega}_{yxx}$ vs $H$ after subtracting the contribution from the $x$ direction. The solid line represents a smooth fit to the data. \subref*{fig:KTO100fluctH_h} Hall voltage fluctuations $\delta V^{2\omega}_{yxx}$ obtained by subtracting the smooth fitting function from the data in \subref*{fig:KTO100fluctH_g}. Measurements were taken at $I^\omega=50\ \mathrm{\mu A}$, $V_\mathrm{G}=0$ V, and $T=2$ K.
}
\end{figure*}

\clearpage
\begin{figure*}
\begin{subcaptiongroup}
\includegraphics[width=.8\columnwidth]{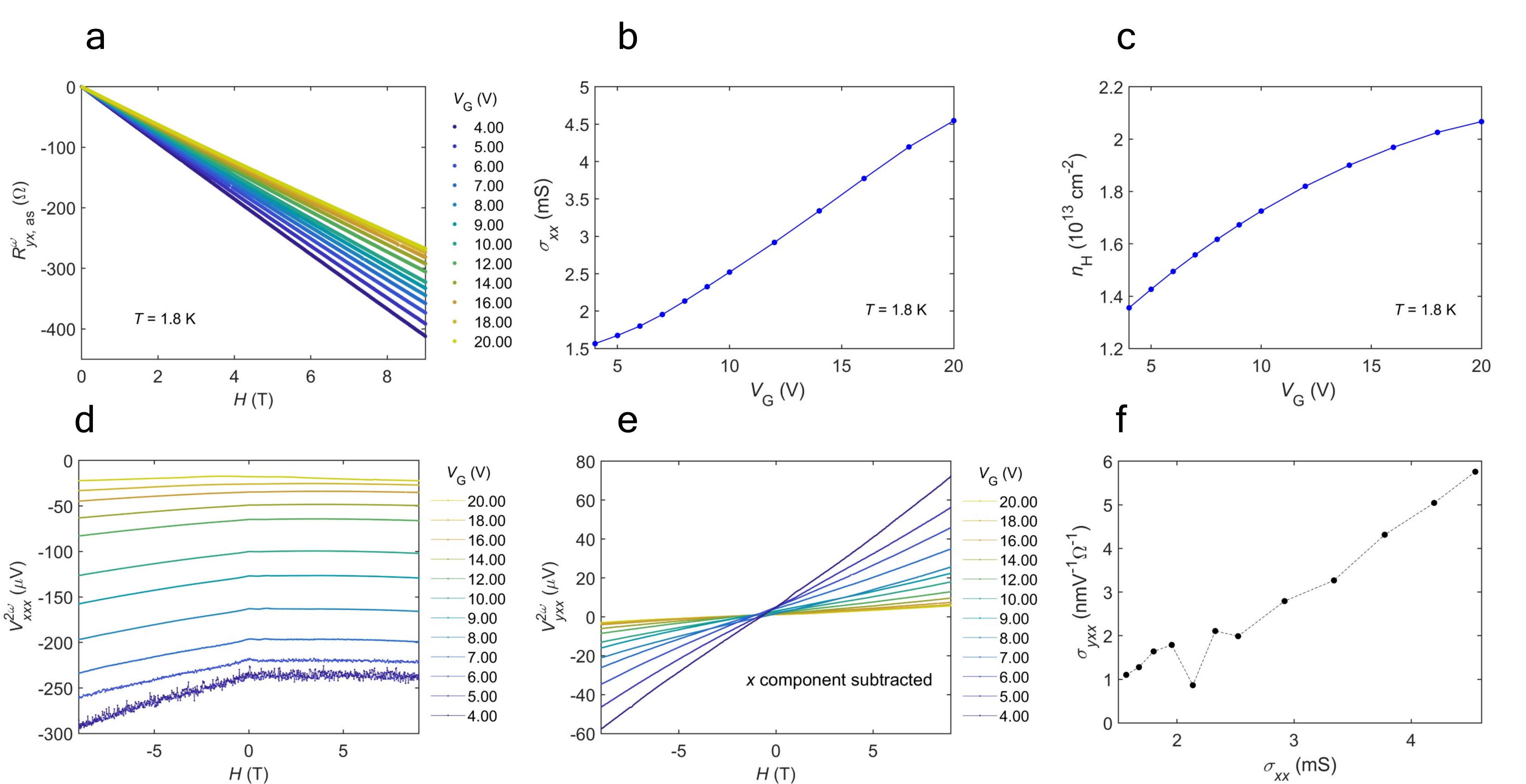}% Here is how to import EPS art
\phantomcaption\label{fig:SLTO111nonlinear_a}
\phantomcaption\label{fig:SLTO111nonlinear_b}
\phantomcaption\label{fig:SLTO111nonlinear_c}
\phantomcaption\label{fig:SLTO111nonlinear_d}
\phantomcaption\label{fig:SLTO111nonlinear_e}
\phantomcaption\label{fig:SLTO111nonlinear_f}
\end{subcaptiongroup}
\captionsetup{subrefformat=parens}
\caption[Nonlinear transport in $\mathrm{LaTiO_3/SrTiO_3}(111)$]{\label{fig:SLTO111transport} \justifying \textbf{Nonlinear transport in $\mathbf{LaTiO_3/SrTiO_3}(111)$}. \subref*{fig:SLTO111nonlinear_a} Antisymmetric part of the Hall resistance, $R^{\omega}_{yx,\,\mathrm{as}}$, as a function of $H$ for various gate voltages $V_{\mathrm{G}}$. 
\subref*{fig:SLTO111nonlinear_b} Longitudinal conductivity $\sigma_{xx}$ as a function of $V_{\mathrm{G}}$. 
\subref*{fig:SLTO111nonlinear_c} Extracted Hall electron density $n_{\mathrm{H}}$ as a function of $V_{\mathrm{G}}$. 
\subref*{fig:SLTO111nonlinear_d} Second-harmonic longitudinal voltage $V^{2\omega}_{xxx}$ as a function of $H$. 
\subref*{fig:SLTO111nonlinear_e} Second-harmonic transverse voltage $V^{2\omega}_{yxx}$ as a function of $H$. 
\subref*{fig:SLTO111nonlinear_f} Nonlinear Hall conductivity $\sigma_{yxx}$ as a function of $\sigma_{xx}$. Measurements were taken at $T=1.8$ K.}
\end{figure*}

    \putbib[Ref_si]
\end{bibunit}

\end{document}